\documentclass[12pt]{scaleai-paper}
\usepackage[square,sort&compress,numbers]{natbib}

\usepackage{mathpazo}
\usepackage{wrapfig}

\usepackage[scaled=0.92]{helvet} % Sligh
\usepackage{fontawesome5}
\usepackage{subcaption}
\usepackage{hyperref}
\usepackage{tcolorbox}
\usepackage{graphicx}
\usepackage{upquote}
\tcbuselibrary{skins,breakable}
\usepackage{listings}
\usepackage{longtable}
\usepackage{booktabs} % For professional looking tables
\usepackage{array}    % For better column formatting
\usepackage{ragged2e} % For text wrapping alignment
\usepackage{algpseudocode}
\usepackage{algorithm}
\usepackage{hyperref}       % hyperlinks
\hypersetup{
    colorlinks=true,
	linkcolor=blue,
	filecolor=magenta,      
	urlcolor=blue,
	citecolor=black,
	pdfinfo={
        Title={Audio MultiChallenge: A Multi-Turn Evaluation of Spoken Dialogue Systems on Natural Human Interaction},
        Subject={},
        Keywords={},
    }
}
\usepackage{longtable}
\usepackage{tabularx}
\usepackage[utf8]{inputenc} % allow utf-8 input
\usepackage[T1]{fontenc}    % use 8-bit T1 fonts
\usepackage{hyperref}       % hyperlinks
\usepackage{url}            % simple URL typesetting
\usepackage{booktabs}       % professional-quality tables
\usepackage{amsfonts}       % blackboard math symbols
\usepackage{nicefrac}       % compact symbols for 1/2, etc.
\usepackage{microtype}      % microtypography
\usepackage{booktabs}
\usepackage{multirow}
\usepackage{amsmath}
\usepackage{xspace}
\usepackage{graphicx}
\usepackage[table]{xcolor}
\usepackage{array}
\usepackage{wrapfig}

\usepackage{enumitem}
\setlist[itemize]{leftmargin=*}
\setlist[enumerate]{leftmargin=*}

\usepackage{graphicx}
\usepackage{tcolorbox}

\usepackage{amsmath}
\usepackage{soul}
\usepackage{cleveref}
\usepackage{listings}
\usepackage{tikz}
\usepackage{eso-pic}

\let\svthefootnote\thefootnote
\newcommand\freefootnote[1]{%
  \let\thefootnote\relax%
  \footnotetext{#1}%
  \let\thefootnote\svthefootnote%
}

\makeatletter
\renewcommand\AB@affilsepx{, \protect\Affilfont}
\makeatother

\title{Audio MultiChallenge: A Multi-Turn Evaluation of Spoken Dialogue Systems on Natural Human Interaction}

\author{Advait Gosai$^*$}
\author{Tyler Vuong$^*$}
\author{Utkarsh Tyagi}
\author{Steven Li}
\author{Wenjia You}
\author{Miheer Bavare}
\author{Arda Uçar} 
\author{Zhongwang Fang}
\author{Brian Jang}
\author{Bing Liu}
\author{Yunzhong He}

\affil{Scale AI}

\newcommand{\authoremail}{%
  \vspace*{-1.5em}\noindent
    \faEnvelope\  \texttt{advait.gosai@scale.com} \quad 
    \faDatabase\  \href{https://huggingface.co/datasets/ScaleAI/audiomc}
    {\texttt{huggingface.co/ScaleAI/audiomc}} \quad 
    \faGlobe\  \  \href{https://scale.com/leaderboard/audiomc}
    {\texttt{scale.com/leaderboard/audiomc}}
}

\newcommand{\benchmarkname}{Audio MultiChallenge\xspace}

\begin{document}
\maketitle
\authoremail
\def\thefootnote{*}\footnotetext{Indicates equal contribution}\def\thefootnote{\arabic{footnote}}
\begin{abstract}
End-to-end (E2E) spoken dialogue systems are increasingly replacing cascaded pipelines for voice-based human-AI interaction, processing raw audio directly without intermediate transcription. Existing benchmarks primarily evaluate these models on synthetic speech and single-turn tasks, leaving realistic multi-turn conversational ability underexplored. We introduce \textbf{\benchmarkname}, an open-source benchmark to evaluate E2E spoken dialogue systems under natural multi-turn interaction patterns. Building on the text-based MultiChallenge framework, which evaluates \textit{Inference Memory}, \textit{Instruction Retention}, and \textit{Self Coherence}, we introduce a new axis \textit{Voice Editing} that tests robustness to mid-utterance speech repairs and backtracking. We further augment each axis to the audio modality, such as introducing Audio-Cue challenges for \textit{Inference Memory} that require recalling ambient sounds and paralinguistic signals beyond semantic content. We curate 452 conversations from 47 speakers with 1,712 instance-specific rubrics through a hybrid audio-native agentic and human-in-the-loop pipeline that exposes model failures at scale while preserving natural disfluencies found in unscripted human speech. Our evaluation of proprietary and open-source models reveals that even frontier models struggle on our benchmark, with Gemini 3 Pro Preview (Thinking), our highest-performing model achieving a 54.65\% pass rate. Error analysis shows that models fail most often on our new axes and that Self Coherence degrades with longer audio context. These failures reflect difficulty of tracking edits, audio cues, and long-range context in natural spoken dialogue. \benchmarkname provides a reproducible testbed to quantify them and drive improvements in audio-native multi-turn interaction capability.

\end{abstract}

\begin{figure}[!h]
     \centering
     \includegraphics[width=\textwidth]{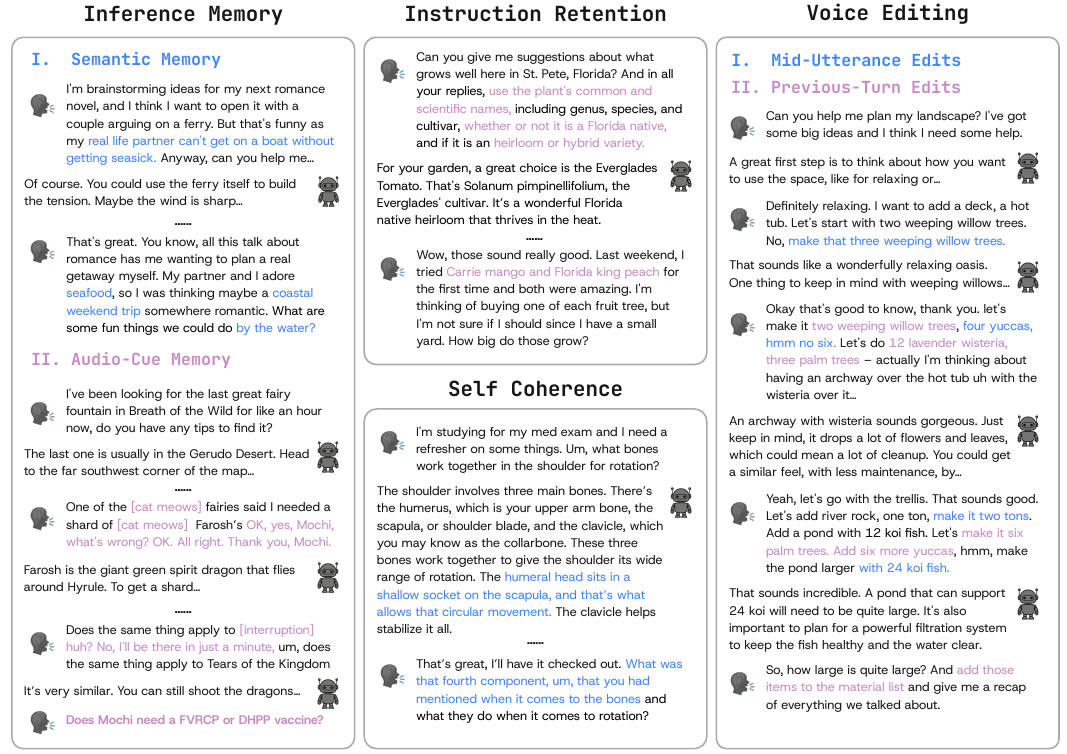}
     \caption{Samples from \benchmarkname. Highlighted text shows key interaction patterns per axis.}
     \label{fig:axes_samples}
     \vspace{-1em}
 \end{figure}

\section{Introduction}

Spoken dialogue systems \cite{Jokinen2010-lw} are evolving from cascaded pipelines comprising Automatic Speech Recognition (ASR) \cite{Radford2022RobustSR}, text-based Large Language Models (LLMs) \cite{brown2020languagemodelsfewshotlearners}, and Text-to-Speech (TTS) systems \cite{wang2017tacotronendtoendspeechsynthesis} to native end-to-end (E2E) architectures that map input speech directly to conversational responses without an intermediate transcription step \cite{nguyen2023generative, arora2025chain, Qwen3-Omni}. Speech-to-speech (S2S) models and broader audio language models (Audio LMs) process raw or minimally processed acoustic features and generate spoken or textual outputs, often operating over shared latent representations or learned discrete speech units rather than relying solely on text tokens \cite{nguyen2023generative,zhang2023speechgptempoweringlargelanguage,borsos2023audiolmlanguagemodelingapproach,kyutai2024moshi,défossez2022highfidelityneuralaudio}. This integration can reduce response latency and mitigate error propagation, while better preserving paralinguistic and acoustic information that is often lost in cascaded systems \cite{cui-etal-2025-recent, ji2024wavchatsurveyspokendialogue}.

%However, these architectural advancements provide little insight into a model’s ability to maintain instructions, reason over long dialogue histories, or remain semantically consistent in multi-turn interactions.

As E2E systems move toward real-world user-facing interaction, their evaluation must reflect the demands of extended, spontaneous dialogue rather than isolated prompt–response pairs or short, synthetic TTS inputs common in existing benchmarks \cite{voicebench, audiobench, bigbench_audio}. Evidence from text-only LLMs shows that strong single-turn performance does not guarantee robust multi-turn behavior \cite{deshpande-etal-2025-multichallenge, laban2025lost, oolong}. These weaknesses are amplified in the audio modality. Audio is temporally continuous and often represented as high-frame-rate token streams, which rapidly consume context and make long-horizon state tracking harder than in text \cite{borsos2023audiolmlanguagemodelingapproach, cui-etal-2025-recent}. Consequently, while existing Audio LM benchmarks suggest strong performance on short, clean, or synthetic inputs, robustness in realistic multi-turn settings remains understudied. Even recent multi-turn efforts are limited to short dialogues \cite{mtalk2025}, leaving unclear whether systems can handle key interaction patterns such as maintaining self-consistency, following user constraints, leveraging audio cues over long contexts, and recovering from user repairs like mid-utterance self-corrections and barge-ins.

To rigorously evaluate these dimensions, we open-source \textbf{\benchmarkname}, an extension of the text-based MultiChallenge evaluation framework \cite{deshpande-etal-2025-multichallenge} to the audio modality through three key contributions:

  \begin{itemize}

    \item \textbf{Audio-native axes and tasks.} We introduce a novel \emph{Voice Editing} axis that evaluates robustness to mid-utterance speech repairs and backtracking, capturing natural human behaviors such as self-corrections that are common in spontaneous dialogue. We also add \emph{Audio-Cue Inference Memory} tasks that require recalling ambient sounds or paralinguistic signals from previous turns, rather than only semantic content, and adapt the remaining \textit{Instruction Retention} and \textit{Self Coherence} axes to better reflect the complexities of voice interactions.
    
    \item \textbf{Hybrid data generation for scalability and diversity.} We introduce a data-generation pipeline that combines multi-agent synthetic component with a human-in-the-loop process to produce our final multi-turn conversation prompts. Rather than providing human contributors with verbatim scripts, we provide high-level \textit{blueprints} and interaction goals derived from a scalable agentic loop designed to elicit diverse failure modes. These blueprints encourage improvisational interactions that surface realistic failures, making our dataset sufficiently adversarial while remaining entirely unscripted and preserving paralinguistics and natural human disfluencies.   

    \item \textbf{Fine-grained rubric-based evaluation.} We replace coarse binary success metrics with \textbf{1,712} atomic, instance-specific \emph{rubrics} across the dataset that decompose each ideal response into critical sub-requirements, and our LLM-as-a-judge method using these rubrics achieves \textbf{93\%} agreement rate with human evaluators. While satisfying all corresponding rubrics is required to pass a task, individual rubrics provide high-resolution attribution of common model failures.

\end{itemize} 

\benchmarkname comprises \textbf{452} multi-turn conversations from \textbf{47} unique speakers, summing up to \textbf{15} hours of user audio recorded without post-processing to retain ambient acoustic variability.

Our analysis of \benchmarkname reveals substantial gaps in current E2E capabilities across frontier models. Even the highest-performing model, Gemini 3 Pro Preview (Thinking), achieves only 54.65\% pass rate, with other model families, especially open-source ones lagging considerably behind. We find that Self Coherence degrades consistently as cumulative user audio duration increases, dropping from 33.3\% on shorter tasks to 20.0\% on tasks with 3--5 minutes of audio. Models also struggle to recall paralinguistic and ambient signals from previous turns, scoring 36.5\% lower on Audio-Cue Inference Memory than on semantic memory tasks. Voice Editing proves to be the most challenging axis, where behaviors such as mid-utterance self-corrections and prior-turn edits frequently trigger failures that are largely absent in cleaner text baselines. Together, our results suggest the need for audio-native training and evaluation that target realistic multi-turn spoken interaction.

\section{\benchmarkname}
\begin{wraptable}{r}{0.35\textwidth}
\vspace{-1em}
\centering
\small
\renewcommand{\arraystretch}{1.05}

\begin{tabular}{@{} l r @{}}
\toprule

\textbf{Conversations} & 452 \\
\quad Inference Memory & 132 \\
\quad Instruction Retention & 120 \\
\quad Self Coherence & 117 \\
\quad Voice Editing & 83 \\
\midrule

\textbf{Turn Count} & 3-8\\
\midrule
\textbf{Rubrics} & 1,712\\
\midrule
\textbf{Speakers} & 47 \\
\textbf{User Audio Duration} & 14.99 h\\
\textbf{Sampling Rate} & 48 kHz \\
\midrule
\textbf{Language} & English \\
\bottomrule
\end{tabular}
\caption{Dataset Statistics}
\vspace{-1em}
\label{tab:dataset_stats}
\end{wraptable}

In \benchmarkname, we focus on four core axes designed to test the multi-turn capabilities E2E spoken dialogue systems: \textbf{Inference Memory, Instruction Retention, Self Coherence, and Voice Editing}. The fourth axis, Voice Editing, is a novel addition specific to the speech domain.  We introduce this axis to address a fundamental difference between modalities. Unlike text inputs where errors are typically backspaced prior to submission, speech is a continuous stream where the editing process is audible. This necessitates that models dynamically filter retracted information to isolate the user's final intent. Collectively, these categories represent practical, real-world conversational hurdles that remain challenging for current frontier models and have historically been evaluated primarily on single-turn tasks. For each category, we release a set of human-recorded, non-scripted test examples curated according to the specific axis definition.

We explicitly distinguish \benchmarkname from text-based MultiChallenge to address the unique requirements of spoken interaction. Although MultiChallenge offers a strong foundation, simply applying TTS to the existing text dataset is insufficient. For instance, many text-based interactions such as writing code or formatting text do not naturally translate to voice. \benchmarkname instead captures the inherent complexity of spontaneous human speech by curating real recordings, which unlike synthetic or scripted data, include natural disfluencies, backtracking, and third-party interruptions or cross-talk. Our approach allows us to evaluate model robustness in terms of semantic understanding as well as against real-world acoustics, varying accents, and complex paralinguistic cues which text-derived benchmarks fail to capture. We define our axes in detail below.

\subsection{Axes}

\paragraph{Inference Memory}
\label{sec:amc_im}
Inference Memory evaluates the model’s capacity to recall and synthesize scattered user-specified details from previous turns that are implicitly required to satisfy the user's final request. Unlike direct queries (e.g., "What did I say earlier?"), these tasks demand that the model proactively reallocate attention to prior context to inform its reasoning. For example, a standard Semantic test might require remembering a stated nut allergy when generating a dessert recipe. \benchmarkname extends this definition to include \textit{Audio-Cue Gated Inference Memory}. In these scenarios, the critical context to be retained is not explicitly spoken, but is embedded in the acoustic environment or paralinguistic signals. An Audio-Cue gated test might feature the background sound of heavy rain in an initial turn, implicitly constraining a later request for an outfit recommendation. This evaluates the model's ability to retain and integrate both semantic content and acoustic cues over a multi-turn conversation.

\paragraph{Instruction Retention}
Instruction Retention measures a model's ability to follow explicit user instructions consistently across an entire multi-turn conversation. In \benchmarkname, we broaden the instruction space beyond simple stylistic or structural constraints, to \emph{stacked} and \emph{conditional} rule sets. These include multi-constraint directives and trigger-based behaviors, for example changing the response format only after a specific phrase is spoken. We also broaden conversation topics here to include more realistic voice applications such as debate and roleplay. This axis evaluates whether an Audio LM can maintain and correctly apply precise, evolving instructions over long dialogue histories, a failure mode that is frequently observed in speech interactions \cite{lu2024desta2, lu2025speech}.

\begin{wrapfigure}{r}{0.6\textwidth}
\vspace{-3em}
  \begin{center}
    \includegraphics[width=0.58\textwidth]{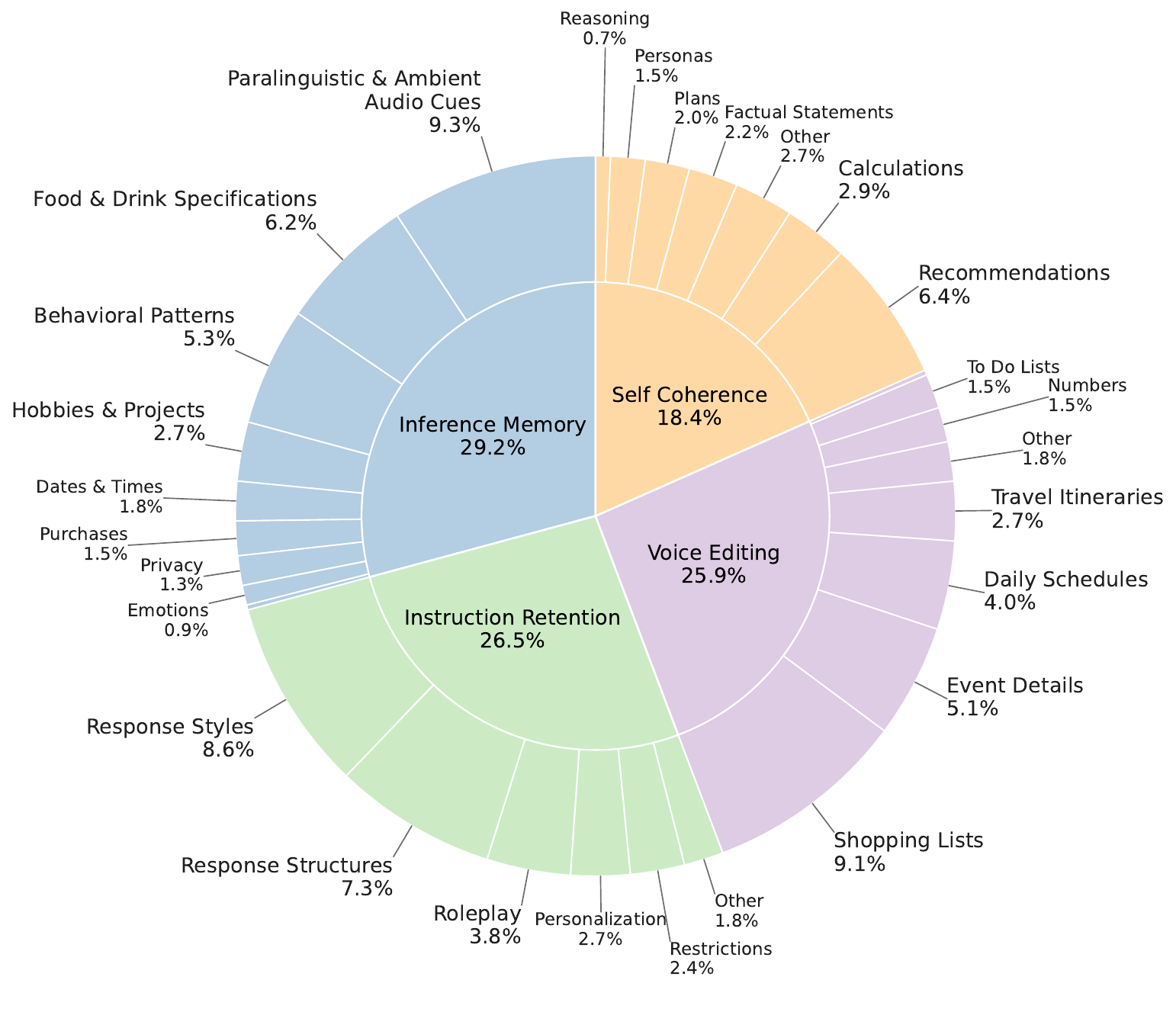}
  \end{center}
  \vspace{-1em}
  \caption{Topic Distribution. Detailed breakdown in Appendix \ref{app:conversation_topics}}
   \vspace{-2em}
\end{wrapfigure}

\paragraph{Self Coherence}

Self Coherence evaluates the model’s ability to maintain internal consistency regarding factual assertions, opinions, and established personas throughout a dialogue. A common failure mode in frontier models is contradicting its prior correct responses to align with a user who questions or challenges them. For \benchmarkname, we refine this category by excluding contrived, adversarial red-teaming prompts in favor of realistic conversational scenarios. While adaptive reasoning is desirable in light of new evidence, we specifically target unwarranted contradictions where the model invalidates its own previous statements without justification.

\paragraph{Voice Editing}
Voice Editing is a novel axis introduced specifically to mimic speech interactions with a personal assistant. We define this axis as the model's ability to recognize and apply immediate, spontaneous speech repairs, such as mid-utterance self-corrections or implemented planned edits that span multiple turns. This axis targets a unique challenge for audio-native models. While LLMs are trained on refined prompts where errors are backspaced or redrafted prior to submission, speech is a continuous stream where the editing process is audible. Consequently, models must dynamically filter out retracted content (e.g., "Let's make it four yuccas, hmm no six.") and resolve implicit contradictions that arise within consecutive turns. Evaluation here focuses on the model's robustness to these natural disfluencies and its capacity to discern the user's final intent amidst the messiness of spoken interaction.

\subsection{Dataset Curation}

\begin{figure}[!ht]
     \centering
     \includegraphics[width=\textwidth]{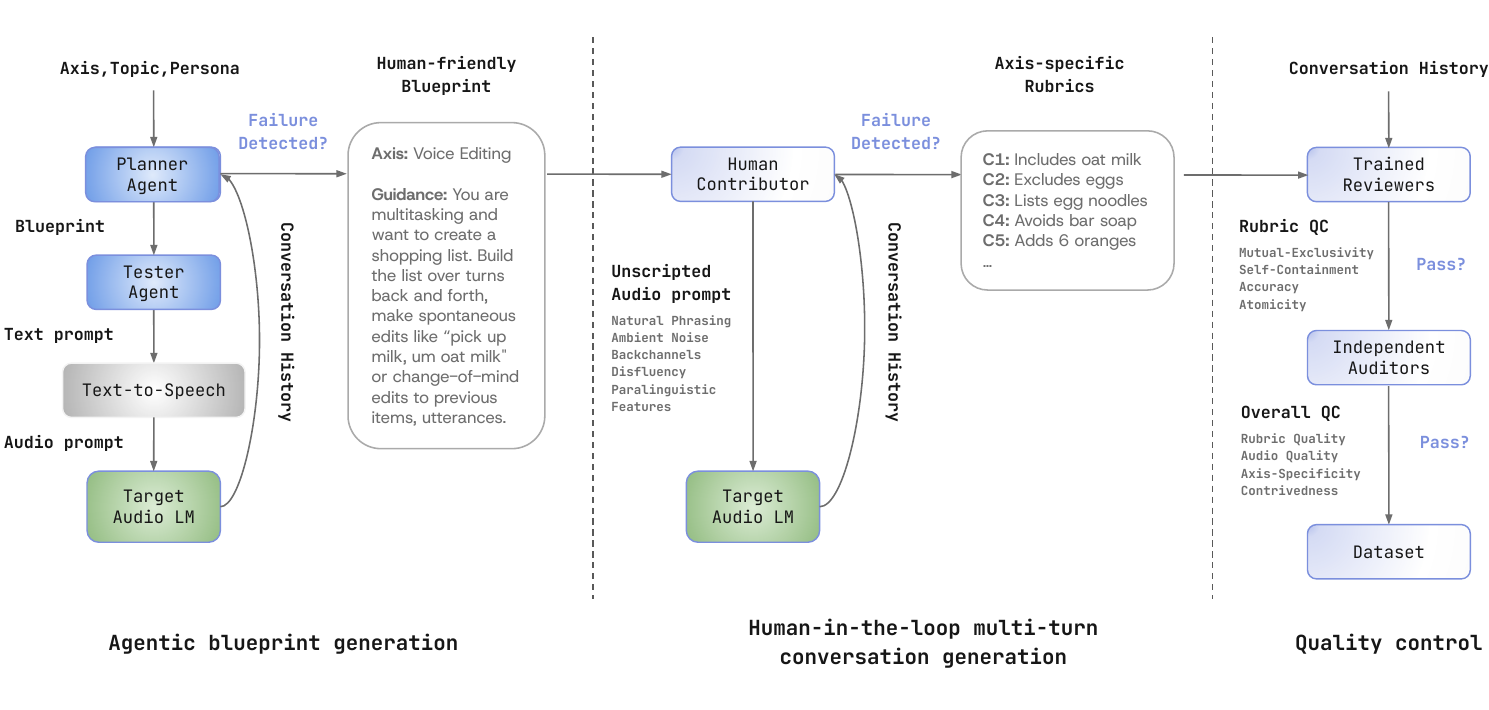}
     \caption{\benchmarkname dataset generation process that utilizes an agentic blueprint generation followed by human-in-the-loop multi-turn conversation generation and quality check. Conversation topic seeds, generated human-friendly blueprints and data samples are provided in the Appendix.}
     \label{fig:data_pipeline}
     \vspace{-1em}
 \end{figure}

Producing \textit{naturalistic, diverse,} and \textit{challenging} test examples for \benchmarkname is resource intensive. We optimize our dataset curation using a two-stage hybrid approach that combines (i) a multi-agent pipeline to synthetically explore Audio LM failure modes at scale with a (ii) human-in-the-loop stage that translates these failure blueprints into realistic conversations and corresponding evaluation rubrics. An overview is shown in Figure \ref{fig:data_pipeline}, and we describe each step below.

\paragraph{Generating Synthetic Audio LM Failures}
We implement an automated multi-agent framework to generate synthetic conversations that elicit failures along a targeted axis. This approach follows the synthetic generation methodology of MultiChallenge, adapted to the audio modality.

As shown in Figure \ref{fig:data_pipeline}, the process begins by sampling a target Evaluation Axis $A$ (e.g. Inference Memory) along with a seed topic $T$ (e.g. DietaryRestrictions) and persona $P$ (e.g. Content creator) from curated taxonomies. A Planner Agent (text LLM) then devises an initial conversation strategy aimed at triggering a specific failure mode while adhering to the sampled topic. Based on this strategy, a Tester Agent (text LLM) generates initial user prompt text, which is converted to speech using TTS to simulate audio input.  This audio is then fed to the Target Agent (Audio LM). The Planner Agent evaluates the Target Agent's output to detect failure. If the model succeeds, the Planner updates the strategy to introduce higher complexity in the subsequent turn. This cycle repeats until a failure is detected or a maximum turn count is reached.  Once a failure is detected, the Planner Agent analyzes the history $H$ and extracts the core logic or trick that caused the model to fail along the axis. This is synthesized into a concise Human-friendly Blueprint $B_{final}$, serving as a high-level strategic guide for human contributors. In this study, we use \texttt{gpt-4o-mini-tts} by OpenAI as the TTS system, randomly sample between GPT 4o Audio Preview and Gemini 2.5 Pro for the Target Agent, and utilize the reasoning capability of OpenAI o3 as the Planner that guides GPT 4o as the User Agent. The full algorithm is provided in Appendix \ref{app:synthetic_loop}. 

\paragraph{Human-In-The-Loop Conversation Generation}

Using axis-specific blueprints as seeds, we collect audio recordings from human contributors. Each contributor attempts to induce a verifiable failure within 3 to 8 turns. Contributors receive the blueprint and a detailed definition of the target axis, but the blueprint is only strategic guidance. They are instructed to rewrite the plan in their own words, introduce new constraints, or pivot topics as needed to expose model weaknesses. This encourages spontaneous, in-the-wild interaction, including pauses, hesitations, rambling, and mid-utterance self-corrections. On average, 65\% of attempted conversations leveraged a synthetic blueprint, highlighting their effectiveness. We provide blueprint samples in Appendix \ref{app:blueprints}.

To collect natural speech data, contributors interact with an Audio LM (either GPT 4o Audio Preview and Gemini 2.5 Pro) in a turn-based manner until a failure is manually detected. For each turn, the contributor records an audio prompt, which is processed by the Target Audio LM. The model output is returned as text for immediate review. After each turn, contributors check that the interaction remains challenging and coherent, avoids premature failures unrelated to the assigned axis, and continues the conversation naturally. If these conditions are met, the contributor proceeds to the next turn. Once an objective failure is triggered under the assigned axis, the conversation ends.

\begin{figure}[!t]
     \centering
     \includegraphics[width=\textwidth]{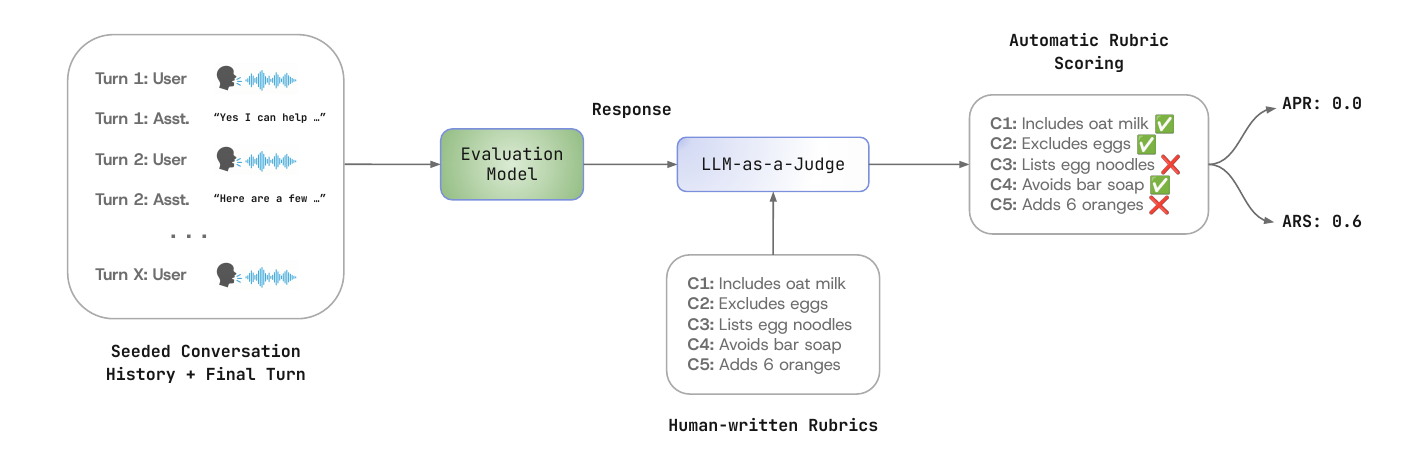}
     \caption{We utilize a fixed-context evaluation protocol where each model receives identical conversation history and is evaluated on its response to the final user turn. An LLM-as-a-judge is subsequently used to evaluate the text stream of the model response against our human-written rubrics. }
     \label{fig:model_input}
     \vspace{-1em}
 \end{figure}

\paragraph{Human-Authored Rubrics for Model Evaluation}

Once the conversation is marked as a failure along the specific axis, contributors formulate a set of atomic, binary criteria specifically designed to evaluate the model response of the final turn where the failure occurred. Rather than creating a single, unstructured ideal rubric, we require contributors to decompose the ideal response into a set of multiple discrete rubrics. To ensure these rubrics strictly measure the targeted axes, we instruct contributors to generate criteria that correspond exclusively to both the assigned axis and the final user turn. General conversational requirements related to only the final user turn or generic axis definitions related to the assigned axis only are explicitly excluded. By breaking down the ideal response into self-contained, binary statements, our approach allows an LLM-as-a-judge to verify individual rubric criterion with higher precision than when scoring against a single reference text.

\paragraph{Quality Control}
We construct \benchmarkname with  highly trained core contributors and a three-stage review process. Each task is audited by specialized reviewers, an independent quality control team, and our internal research team. We evaluate quality along conversation naturalness, failure validity, prompt diversity, audio quality, and rubric quality, as summarized in Figure \ref{fig:data_pipeline}.

\section{Evaluation}

\paragraph{Fixed-Context Evaluation Protocol}
To ensure a standardized evaluation, we adopt a fixed-context protocol, as depicted in Figure \ref{fig:model_input}. Specifically, for each test example, we pre-seed the full conversation history (both user and assistant turns) and evaluate only the model’s response to the final turn using our LLM-as-a-judge approach. User turns are provided in context as base64-encoded audio files, while assistant turns are provided as raw text. This contrasts with sequential evaluation methods used in prior work, where a model generates responses turn-by-turn to pre-recorded user prompts \cite{mtalk2025}. We avoid the sequential approach because it can suffer from conversation drift: if a model’s response in the first turn diverges even slightly from the reference transcript, subsequent pre-recorded user follow-ups and their associated rubrics may become semantically invalid. Our approach encapsulates more natural back-and-forth interaction patterns and ensures that every model receives an identical context state, enabling a fair, reproducible comparison.

\paragraph{Evaluation Metrics} We utilize the Average Pass Rate (APR) metric \cite{guo2025seeingevaluatingmultimodalllms} to evaluate model performance on \benchmarkname. Concretely, a model’s response is considered successful if and only if it satisfies all corresponding rubrics; otherwise, it is marked as a failure. We report the Overall APR as the percentage of tasks successfully completed across the dataset. This methodology aligns with MultiChallenge \cite{deshpande-etal-2025-multichallenge}, where a single, binary weighted target question was defined to grade each tasks. Instead of restricting our data to one rubric criteria, we allow for multiple binary weighted atomic rubrics for increased LLM judge reliability, and equivalently calculate model performance as their APR. Additionally, we report an Average Rubric Score (ARS) \cite{guo2025seeingevaluatingmultimodalllms} or a \textit{rubric compliance} \cite{sharma2025researchrubricsbenchmarkpromptsrubrics} score for each model  which represents the average percentage of satisfied rubrics per task. We define our evaluation metrics formally in Appendix \ref{app:eval_metrics}.

\definecolor{highgreen}{HTML}{D1E7DD} 
\definecolor{highblue}{HTML}{CFE2FF}
\definecolor{aprapproach}{gray}{0.9}

\newcommand{\cgreen}{\cellcolor{highgreen}}
\newcommand{\cblue}{\cellcolor{highblue}}

\begin{table}[!ht]
\centering
\resizebox{0.95\textwidth}{!}{%
\begin{tabular}{lc cc | cccc}
\toprule
\multirow{2}{*}{\textbf{Model}} & \multirow{2}{*}{\textbf{Output Modality}} & \multicolumn{2}{c}{\textbf{Overall}} & \multicolumn{4}{c}{\textbf{Per-Axis APR (\%)}} \\
\cmidrule(lr){3-4} \cmidrule(lr){5-8}
 & & \textbf{APR (\%)} & \textbf{ARS (\%)} & \textbf{IM} & \textbf{IR} & \textbf{SC} & \textbf{VE} \\
\midrule

% Group: Gemini Models
Gemini 3 Pro Preview (Thinking) & Text & \cgreen 54.65 & \cgreen 82.54 & \cgreen 57.58 & \cgreen 65.283 & 40.96 & \cgreen 49.57 \\
Gemini 2.5 Pro (Thinking) \cite{comanici2025gemini} & Text & 46.90 & 78.91 & 44.70 & 56.67 & 40.96 & 43.59 \\
Gemini 2.5 Flash (Thinking) & Text & 40.04 & 74.59 & 30.30 & 37.50 & \cgreen 54.22 & 43.59 \\
Gemini 2.5 Flash & Text & 26.11 & 65.42 & 19.70 & 29.17 & 31.33 & 26.50 \\
\addlinespace

% Group: GPT Realtime
GPT Realtime\footnotemark[1] & Text & 23.45 & 65.95 & 21.97 & 26.67 & 22.89 & 22.22 \\
GPT Realtime & Audio & 20.35 & 63.03 & 19.70 & 20.00 & 26.51 & 17.09 \\
\addlinespace

% Group: GPT-4o Audio Preview
GPT 4o Audio Preview \cite{hurst2024gpt} & Text & 25.44 & 67.58 & 23.48 & 27.50 & 31.33 & 21.37 \\
GPT 4o Audio Preview & Audio & 23.23 & \cblue 64.25 & \cblue 21.21 & \cblue 30.83 & 24.10 & 17.09 \\
\addlinespace

% Group: GPT-4o Mini Audio Preview
GPT 4o Mini Audio Preview & Text & 14.82 & 57.01 & 15.15 & 16.67 & 20.48 & 8.55 \\
GPT 4o Mini Audio Preview & Audio & 13.05 & 54.91 & 9.85 & 12.50 & 21.69 & 11.11 \\
\addlinespace

Qwen 3 Omni \cite{Qwen3-Omni} & Audio & \cblue 24.34 & 63.43 & 20.45 & 27.50 & \cblue 30.12 & \cblue 21.37\\
Qwen 2.5 Omni \cite{xu2025qwen2} & Text & 11.95 & 45.39 & 12.88 & 13.33 & 18.07 & 5.13 \\
\addlinespace

% Group: Voxtral
Voxtral Small 24B \cite{liu2025voxtral} & Text & 26.33 & 66.59 & 20.45 & 25.00 & 27.71 & 33.33 \\
\addlinespace

% Group: Mimo
Mimo Audio 7B Instruct (Thinking) \cite{coreteam2025mimoaudio} \footnotemark[1]& Text & 19.69 & 55.78 & 19.70 & 22.50 & 27.71 & 11.11 \\
Mimo Audio 7B Instruct & Text & 18.58 & 53.86 & 17.42 & 20.00 & 27.71 & 11.97 \\
\addlinespace

% Group: Phi
Phi 4 MM Instruct \cite{abouelenin2025phi} & Text & 15.49 & 46.96 & 16.67 & 16.67 & 28.92 & 3.42 \\
\addlinespace

% Group: Gemma
Gemma 3n E4B IT \cite{gemma_3n_2025} & Text & 15.49 & 43.11 & 14.39 & 22.50 & 26.51 & 1.71 \\
\addlinespace

% Group: Other
Kimi Audio 7B Instruct \cite{ding2025kimi} & Text & 13.72 & 48.36 & 14.39 & 16.67 & 20.48 & 5.13 \\
Kimi Audio 7B Instruct & Audio & 10.40 & 22.08 & 15.91 & 6.67 & 13.25 & 5.98 \\
\addlinespace

LFM2 Audio 1.5B \cite{amini2025lfm2} & Audio & 9.29 & 19.10 & 15.15 & 7.50 & 15.66 & 0.00 \\

\bottomrule
\end{tabular}%
}
\caption{Overall and Per-Axis scores on \benchmarkname per model and output modality. APR serves as final scores. IM: Inference Memory, IR: Instruction Retention, SC: Self Coherence, VE: Voice Editing. Green highlights indicate best performance with output modality as Text; blue highlights indicate best Audio output performance.}
\vspace{-1em}
\label{tab:results}
\end{table}
\footnotetext[1]{These models previously report scores on a TTS set of MultiChallenge (referring to it as MultiChallenge Audio) which we distinguish from.}

\subsection{Results}

We present scores on \benchmarkname across a variety of E2E architectures in Table \ref{tab:results}, along with their performance per-axis. We observe that Gemini 3 Pro Preview (Thinking) achieves the highest overall score of 54.65\%, followed by its counterparts Gemini 2.5 Pro (Thinking) and 2.5 Flash (Thinking). In comparison, the remainder of proprietary and open-source models perform modestly, with GPT 4o Audio Preview and Voxtral Small 24B leading, scoring a 25.44\% and 26.33\% APR respectively, closely followed by GPT Realtime with an APR of 23.45\% (output modality set to Text). Open-source S2S architectures in particular struggle the most, with no model surpassing Qwen 3 Omni which achieves a modest 24.34\% APR. Analyzing per-axis scores, our newly adapted Voice Editing and Inference Memory prove to be the most challenging tasks for frontier models, with them scoring 17.99\% and 21.55\% on average, respectively. We highlight the following key findings and provide samples of model failures in Appendix \ref{app:model_failures}.

\paragraph{Performance varies over output modalities} We observe a "modality gap" when evaluating models that support both text and audio output streams. Under a Text only output configuration, these models score 19.36\% on average, however, when the same models are configured to output Audio (or joint Audio and Text depending on their architecture), their performance degrades to 16.76\%. These findings are independently reported in other recent works \cite{coreteam2025mimoaudio}, highlighting the need for improved S2S post-training data to overcome this gap in performance.

\begin{figure}[!t]
    \centering
    \includegraphics[width=\textwidth]{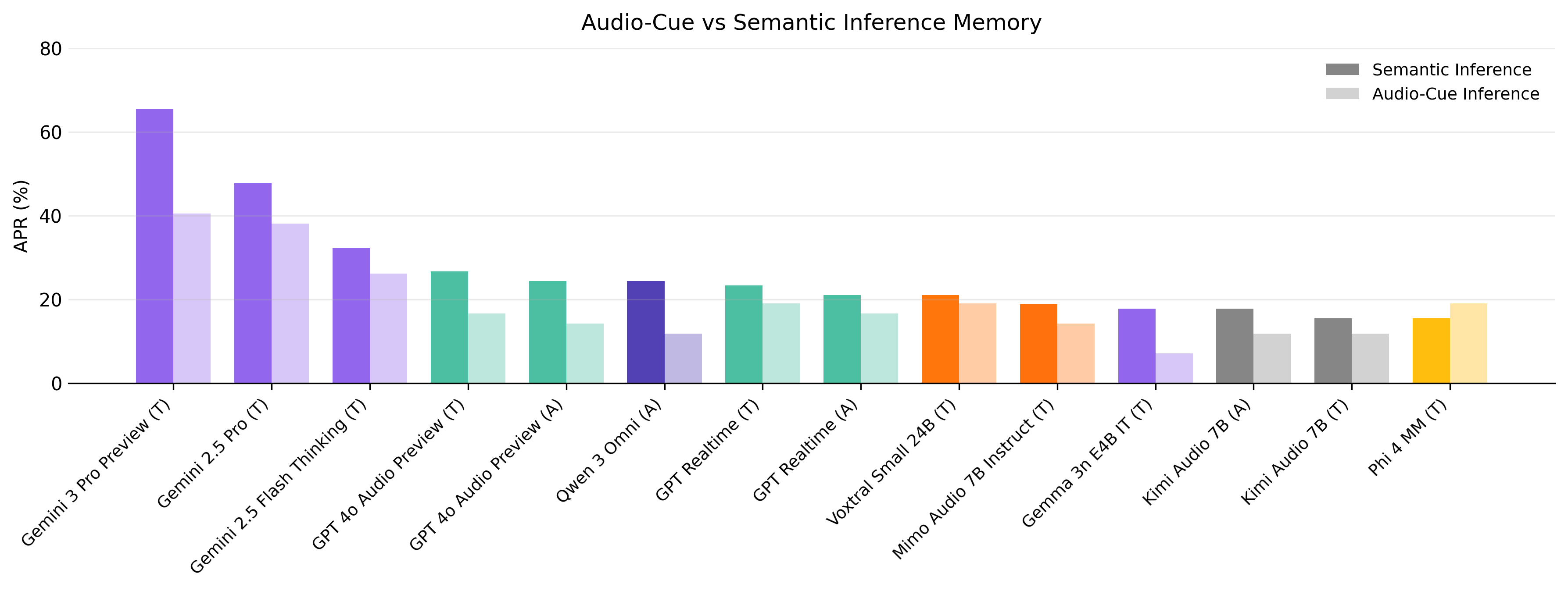}
    \caption{APR (\%) scores on Audio-Cue vs. Semantic Inference Memory tasks. (T) indicates text output, (A) indicates audio output.}
    \label{fig:audio-im}
\end{figure}

\paragraph{Audio-Cue Inference Memory lags substantially} We classify Audio-Cue gated Inference Memory tasks ($n=42$) as ones where the information the model must remember from a previous turn is contained only in the audio signal rather than in the words spoken, as described in Section \ref{sec:amc_im}. Our examples span both \emph{paralinguistic} cues such as the speaker's tone and \emph{ambient} cues such as background sound. Figure \ref{fig:audio-im} shows a significant gap in remembering these audio cues in our multi-turn setup. Across most models, performance drops substantially when the task requires inferring paralinguistic features or ambient sounds from previous turns, with ARS falling by 36.5\% relatively compared to semantic memory on average across models. This unique subset of our data highlights the limitations of current E2E spoken dialogue systems when it comes to audio-native state tracking and recall, which is fundamental to realtime and multi-turn voice conversations.

\begin{figure}[!t]
    \centering
    \vspace{-0.5em}
    \includegraphics[width=\textwidth]{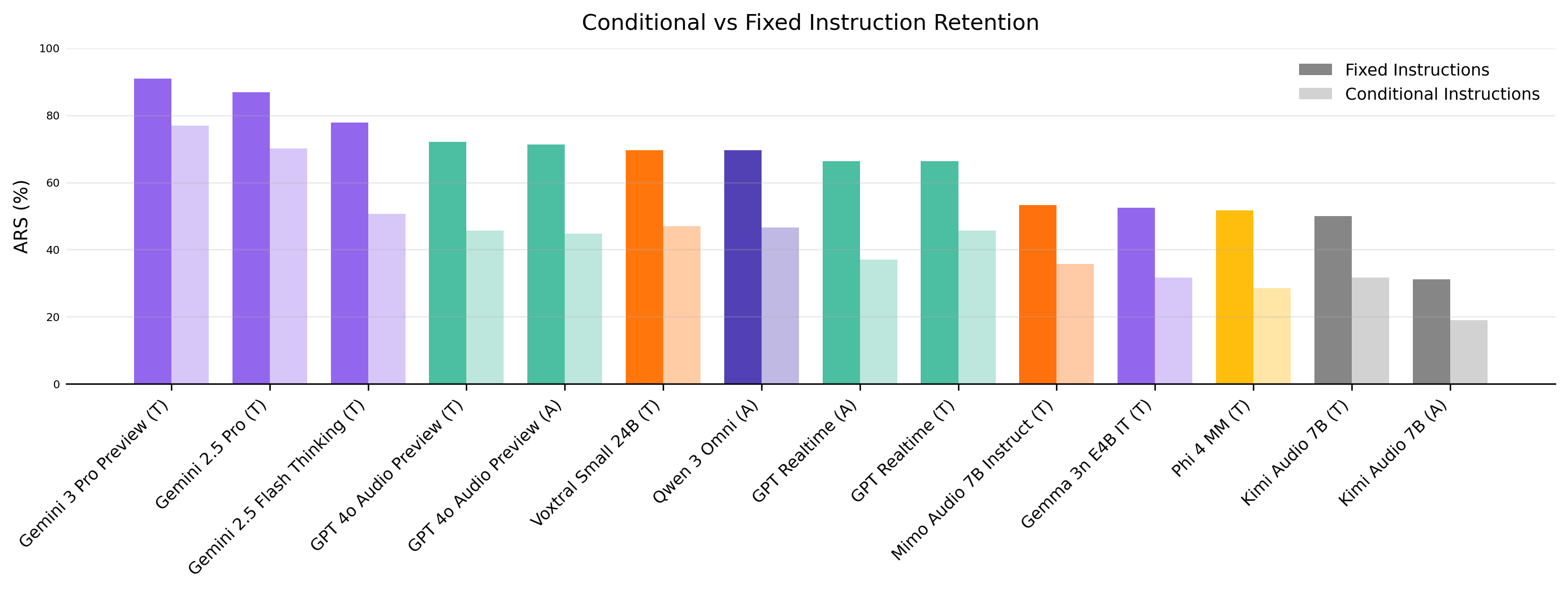}
    \caption{ARS (\%) on Instruction Retention tasks classified by rubric conditionality. ARS is reported instead of APR since the classification is at the rubric level -- whereas a single task can have both conditional and fixed rubrics. (T) indicates text output, (A) indicates audio output.}
    \label{fig:cond-ir}
    \vspace{-1.5em}
\end{figure}

\paragraph{Conditional Instructions cause model failures} Frontier models score 25.08\% APR and 46.11\% ARS on average on Instruction Retention (IR) tasks. To analyze the main sources of error, we automatically cluster each IR rubric item into two instruction types: \textit{Fixed}, which are unconditional, persistent constraints meant to hold throughout the conversation (e.g., response style or roleplay prompts), and \textit{Conditional}, which are context-dependent instructions that only apply in specific future-turn scenarios. Figure \ref{fig:cond-ir} shows that each model, regardless of their overall performance, demonstrates worse performance in conditional IR. These results support treating conditional instruction following as a distinct, harder sub-problem within IR, and they motivate future work on reliably retrieving and applying conditional rules rather than optimizing only for conventional instruction following.

\paragraph{Self Coherence linearly declines on tasks with increased context} Since the models we benchmark use different audio tokenizers, raw token counts are not directly comparable across systems. We therefore use cumulative user-audio duration per task as a proxy for audio context length. Self Coherence degrades sharply as cumulative user audio increases. Figure \ref{fig:scores-turn-count}b shows that APR on Self Coherence drops from 33.3\% for tasks with 0–60 seconds of cumulative user audio to 20.0\% for tasks with 3–5 minutes, and further to 13.3\% for tasks exceeding 5 minutes (though this final bin contains only 3 samples) when averaged across all models. We analyze duration effects on other axes in Section \ref{sec:turn_count}.

\begin{figure}[!t]
    % \vspace{-2em}
    \centering
    \includegraphics[width=\textwidth]{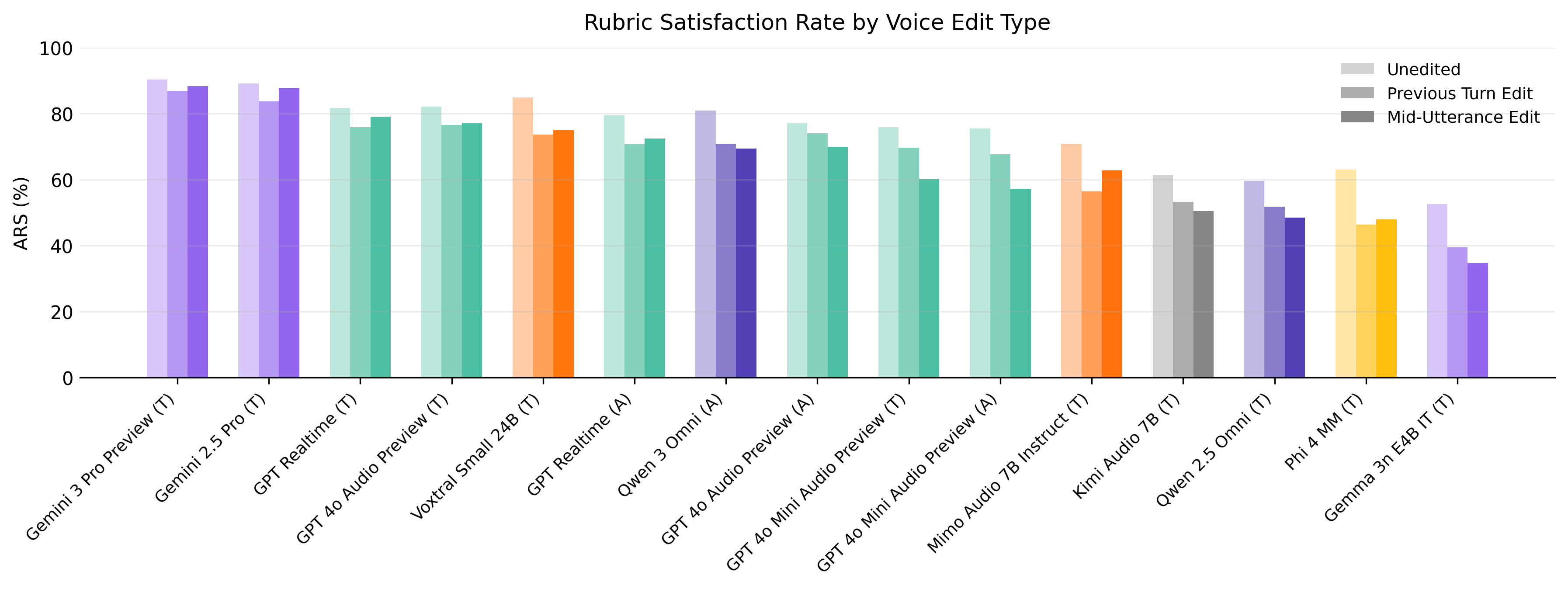}
    \caption{ARS (\%) on Voice Editing tasks classified by the edit type made by the user for each rubric. Unedited refers to statements that the user did not backtrack or revise either within the same turn or in future turns. (T) indicates text output, (A) indicates audio output.}
    \label{fig:ve-rubric-scores}
    \vspace{-1em}
\end{figure}

\paragraph{Mid-Utterance and Prior-Turn Voice Edits hurt performance} Voice Editing remains our most challenging axis, with an average APR of 17.99\%. Most failures on Voice Editing occur when users introduce multiple \textit{mid-utterance} (in-turn speech repairs) and \textit{prior-turn} edits throughout a conversation, with the final prompt typically requesting a summary or revision that integrates all previously specified changes. Successfully completing these tasks requires strong state tracking and multi-constraint integration at the semantic level, as well as robustness to user backchannels and disfluencies at the acoustic level. This helps explain why stronger or reasoning-heavy models can be similarly robust to mid-utterance and prior-turn edits and to unedited rubric items corresponding to simple instructions that were never modified, as shown in Figure \ref{fig:ve-rubric-scores}. Our ablation studies in Section \ref{sec:ablations} further highlight the difficulty of this axis.

\subsection{Analysis}
\label{sec:further_analysis}

\begin{figure}[!h]
    % \vspace{-2em}
    \centering
    \includegraphics[width=\textwidth]{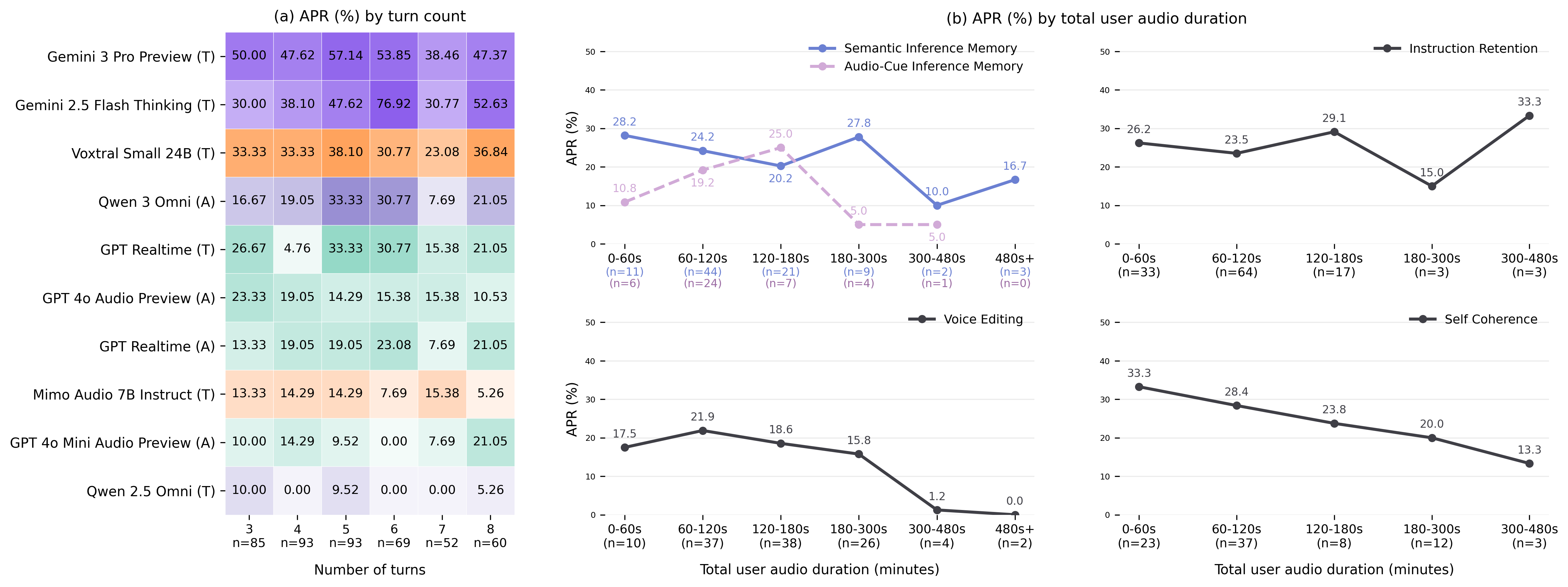}
    \caption{APR (\%) across (a) number of turns, (b) binned total user audio duration (in seconds)}
    \label{fig:scores-turn-count}
    \vspace{-2em}
\end{figure}

\paragraph{Turn Count}
\label{sec:turn_count}
Figure \ref{fig:scores-turn-count}a shows no meaningful relationship between turn count and performance: the scores in \benchmarkname remain largely stable as the number of turns increases. Although the benchmark is multi-turn, the conversations are short relative to the context windows of modern models, so adding turns does not mechanically increase difficulty. This mirrors findings from prior work using fixed-context evaluation protocols \cite{deshpande-etal-2025-multichallenge}. Since each turn's length is also arbitrarily defined, it is a weak measure for conversation histories that models must parse through. Furthermore, since the history for all previous turns is provided to the model and is audited (i.e., free of assistant hallucinations or errors), models can often leverage additional in-context examples at higher turn counts to respond consistently. We observe similarly weak trends when analyzing performance by axis.

\paragraph{Audio Duration}
\label{sec:audio_duration}
Unlike turn count, total user-audio duration is a more direct proxy for effective context length. Figure \ref{fig:scores-turn-count}b shows a clear degradation in Self Coherence as audio duration increases, consistent with the trend discussed above. Voice Editing also worsens with longer histories as average APR falls from 17.5\% in the 3–5 minute bin to 1.2\% and 0\% for 5-8 minute and 8+ minute tasks respectively (though these bins have low sample sizes $n=4, 2$). Inference Memory declines more modestly with duration, driven primarily by Semantic Inference Memory, where longer contexts make targeted recall harder. By contrast, Audio-Cue Inference Memory shows no clear trend with duration; qualitatively, we observe performance is often bottlenecked first by audio perception (e.g., reliably identifying paralinguistic cues or background sounds) rather than retention per se, leading to failures even on short clips. Interestingly, Instruction Retention remains steady or even improves on longer audio tasks because many instruction-following tasks introduce conditional or fixed constraints early in the dialogue that must be followed throughout. Under our fixed-context protocol, the model also receives audited prior assistant turns that already follow these constraints, providing additional in-context demonstrations it can imitate in later turns, consistent with in-context learning effects observed on instruction following in LLMs \cite{dong2024surveyincontextlearning}. 

\begin{wrapfigure}{r}{0.6\textwidth}
\vspace{-2em}
  \begin{center}
    \includegraphics[width=0.58\textwidth]{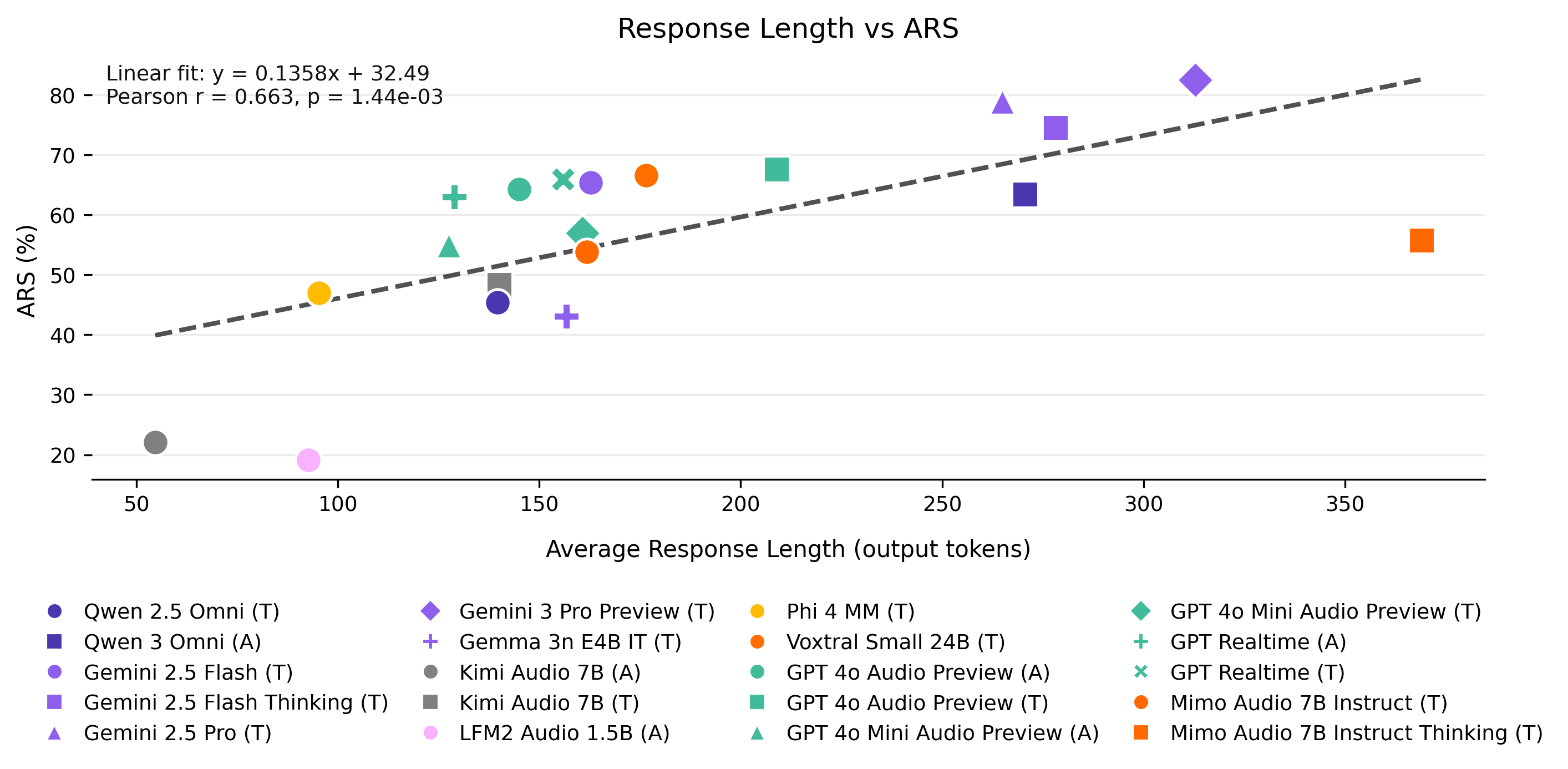}
  \end{center}
     \vspace{-1em}
    \caption{Rubric compliance (ARS) as a function of the model's average response length in tokens.}
    \vspace{-2em}
    \label{fig:response-len}
\end{wrapfigure}

\paragraph{Response Length}
One hypothesis for why models score higher on rubric-based benchmarks is their scores may correlate with response length \cite{sharma2025researchrubricsbenchmarkpromptsrubrics}. Since rubrics evaluate against an explicit list of criteria, an overly verbose model could satisfy more rubric items even if the output is not realistic or human-desirable. A common strategy to mitigate this is to include negatively weighted criteria for incorrect or undesirable content. For \benchmarkname, we adapt this approach by utilizing exclusion rubrics (e.g., “Avoids mentioning leather products”) that explicitly check whether the model omits information the user instructed not to mention, later edited away, or signaled a preference to avoid. Figure \ref{fig:response-len} plots rubric compliance (ARS) against each model’s average response length. Response length is measured by tokenizing each model’s text output stream with the \texttt{tiktoken} library. Despite these controls, we observe a strong, approximately linear relationship between response length and ARS. To assess whether this reflects score hacking, we independently review responses from the Gemini family and find no evidence that their higher scores are driven primarily by verbosity. Instead, their outputs more consistently reflect stronger audio and semantic understanding across axes and more reliable state tracking.

\subsection{LLM-as-a-Judge}
\label{sec:ira}

\begin{wraptable}{r}{0.45\textwidth}
\vspace{-1em}
\centering
\renewcommand{\arraystretch}{1.1}
\begin{tabular}{lcc}
\toprule
\textbf{Judge} & \textbf{Cohen's $\kappa$} & \textbf{Macro F1} \\
\midrule
o4 Mini & 0.873 & 0.937 \\
GPT 5 Mini & 0.870 & 0.935 \\
GPT 4.1 & 0.811 & 0.906 \\
Claude Haiku 4.5 & 0.808 & 0.904 \\
Mistral Medium & 0.787 & 0.894 \\
DeepSeek V3.1 & 0.765 & 0.882 \\
\bottomrule
\end{tabular}
\caption{LLM Judge agreement with humans.}
\label{tab:ira}
\vspace{-2em}
\end{wraptable}

To evaluate our LLM-as-a-judge rubric-grading setup, we measure inter-rater agreement between model-generated and human-assigned labels across the full dataset. Human annotators are instructed to grade randomly sampled GPT 4o Audio Preview or Gemini 2.5 Pro responses according to their corresponding written rubrics.

We aggregate these rubric grades into a set of 1,712 observations and compute Cohen’s $\kappa$ and Macro F1 scores calculated following Arora et al. \cite{arora2025healthbenchevaluatinglargelanguage}. To mitigate the self-enhancement bias known to affect LLM judges \cite{zheng2023judgingllmasajudgemtbenchchatbot, wataoka2025selfpreferencebiasllmasajudge}, we exclude our evaluated ALMs which are capable of text-based prompts from judging. Our results, presented in Table~\ref{tab:ira}, show that all judges achieve extremely high agreement with human graders, indicating the effectiveness of our rubric creation and overall LLM-as-a-Judge setup. Among the tested judges, o4 Mini attains the highest performance, with a Macro F1 score of 0.937, and is therefore selected as our primary LLM judge. We provide our LLM-as-a-Judge setup in Appendix \ref{app:judge_prompt}.

\subsection{Ablation Studies} 
\label{sec:ablations}

\paragraph{TTS vs. Human User Audio} To isolate the impact of spontaneous and disfluent speech, we evaluate our models on synthetic versions of our human user audio, generated by first performing ASR using \texttt{whisper-large-v3} followed by TTS using \texttt{gpt-4o-mini-tts}. We exclude Audio-Cue IM and tasks that contain interruptions or sidebar conversations, resulting in a filtered subset of 385 samples for this study. Results in Table \ref{tab:human-tts-comparison}a depict models configured for text outputs achieve a 7.5\% relative performance improvement when using TTS speech. In contrast, the same architectures configured for audio outputs exhibit a 2.5\% relative decline, suggesting their post-training may be optimized for real human input. 

Table \ref{tab:human-tts-comparison}b shows that Voice Editing benefits the most from TTS under text output configurations. We attribute this effect to the inability of ASR and TTS to fully capture human-like hesitations and unintelligible segments in highly spontaneous speech. The resulting input is slower and more structured, making Voice Editing tasks easier to follow. However, the performance degradation observed for audio output configurations on Voice Editing and Instruction Retention when using TTS requires further investigation. Finally, while this study applies TTS to transcripts derived from unscripted human speech, our internal findings indicate that TTS inputs boost ALM performance even strongly for scripted or fully synthetic dialogue, underscoring the importance of collecting real human recordings to accurately evaluate model performance. 

% Requires (at least): \usepackage{booktabs,graphicx,xcolor,colortbl,subcaption}

\definecolor{pastelgreen}{HTML}{D1E7DD}
\definecolor{pastelred}{HTML}{F8D7DA}
\newcommand{\cpos}{\cellcolor{pastelgreen}}
\newcommand{\cneg}{\cellcolor{pastelred}}

\begin{table}[!h]
\centering

\begin{subtable}[t]{0.49\textwidth}
\centering
\resizebox{\textwidth}{!}{%
\begin{tabular}{lcccc}
\toprule
\textbf{Model} &
\textbf{\shortstack{Output\\Modality}} &
\textbf{\shortstack{Human\\APR (\%)}} &
\textbf{\shortstack{TTS\\APR (\%)}} &
\textbf{\shortstack{$\Delta$ Relative\\(\%)}} \\
\midrule
Gemini 3 Pro Preview & Text & 55.58 & 56.62 & \cpos +1.9 \\
Gemini 2.5 Pro & Text & 47.53 & 49.09 & \cpos +3.3 \\
Gemini 2.5 Flash Thinking & Text & 41.56 & 42.86 & \cpos +3.1 \\
Gemini 2.5 Flash & Text & 25.97 & 27.79 & \cpos +7.0 \\
\addlinespace
GPT-4o Audio Preview & Text & 27.27 & 28.05 & \cpos +2.9 \\
GPT-4o Audio Preview & Audio & 24.42 & 23.64 & \cneg -3.2 \\
\addlinespace
GPT-4o Mini Audio Preview & Text & 15.84 & 17.92 & \cpos +13.1 \\
GPT-4o Mini Audio Preview & Audio & 14.03 & 13.77 & \cneg -1.9 \\
\addlinespace
GPT Realtime & Text & 24.16 & 29.35 & \cpos +21.5 \\
GPT Realtime & Audio & 20.52 & 20.00 & \cneg -2.5 \\
\bottomrule
\end{tabular}%
}
\caption{APR (\%) on Human vs. TTS user audio per model.}
\end{subtable}
\hfill
\begin{subtable}[t]{0.49\textwidth}
\centering
\resizebox{\textwidth}{!}{%
\begin{tabular}{lcccc}
\toprule
\textbf{Axis} &
\textbf{\shortstack{Output\\Modality}} &
\textbf{\shortstack{Human\\APR (\%)}} &
\textbf{\shortstack{TTS\\APR (\%)}} &
\textbf{\shortstack{$\Delta$ Relative\\(\%)}} \\
\midrule
Inference Memory & Text  & 34.05 & 34.88 & \cpos +2.4 \\
Inference Memory & Audio & 19.38 & 20.54 & \cpos +6.0 \\
\addlinespace
Instruction Retention & Text  & 37.12 & 38.95 & \cpos +4.9 \\
Instruction Retention & Audio & 20.80 & 18.52 & \cneg -11.0 \\
\addlinespace
Self Coherence & Text  & 34.18 & 35.62 & \cpos +4.2 \\
Self Coherence & Audio & 24.47 & 26.58 & \cpos +8.6 \\
\addlinespace
Voice Editing & Text  & 30.24 & 33.70 & \cpos +11.5 \\
Voice Editing & Audio & 14.89 & 12.94 & \cneg -13.0 \\
\bottomrule
\end{tabular}%
}
\caption{APR (\%) on Human vs. TTS user audio per axis.}
\end{subtable}
\vspace{-1em}
\caption{Ablation study comparing APR (\%) for select models on TTS vs. Human user audios.}
\label{tab:human-tts-comparison}
\vspace{-1em}
\end{table}

\paragraph{E2E vs. Cascaded Systems} We benchmark frontier LLMs in a cascaded pipeline (ASR + LLM) to test their performance on our dataset.  For all cascaded experiments, we use the same transcripts obtained from \texttt{whisper-large-v3} as the text input to the LLMs.  Table \ref{tab:e2e-vs-cascaded}a reports APR (\%) on our previously filtered subset that excludes audio-cue and interruption tasks. We observe cascaded systems using GPT 5 and Claude Opus 4.5 perform strongly relative to most E2E baselines, though they do not surpass all Gemini models. Notably, cascading reduces performance for Gemini 3 Pro Preview and Gemini 2.5 Pro, with relative drops of 10.3\% and 4.9\%, while Gemini 2.5 Flash Thinking improves by 10.0\%. A possible hypothesis is that newer Gemini models are post-trained more extensively on audio-native data to function as E2E multimodal architectures, rather than solely being strong LLMs.

Table \ref{tab:e2e-vs-cascaded}b reports performance on the excluded audio specific subset containing audio cues or interruptions ($n=67$). As expected, cascaded performance drops sharply for Gemini 3 Pro Preview and Gemini 2.5 Pro, with relative declines of 30.3\% and 24.1\%. In contrast, Gemini 2.5 Flash Thinking shows no change between E2E and cascaded on this subset. While GPT 5, Claude Opus 4.5, and GPT 4o cascaded systems also drop in APR on this subset, they still achieve nontrivial scores. This pattern is consistent with recent findings that audio understanding can improve through text-only post training \cite{rouditchenko2025omnir1reallyneedaudio}, and suggests that reasoners may partially infer some audio cues, particularly paralinguistic and emotional context, from transcripts and surrounding dialogue.

\definecolor{pastelgreen}{HTML}{D1E7DD}
\definecolor{pastelred}{HTML}{F8D7DA}
\definecolor{cascadegray}{gray}{0.92}

\newcommand{\ccascade}{\cellcolor{cascadegray}}

\begin{table}[!h]
\centering

\begin{subtable}[t]{0.48\textwidth}
\centering
\resizebox{\textwidth}{!}{%
\begin{tabular}{lcccc}
\toprule
\textbf{Model} &
\textbf{\shortstack{E2E\\APR (\%)}} &
\textbf{\shortstack{Cascaded\\APR (\%)}} &
\textbf{\shortstack{$\Delta$ Relative\\(\%)}} \\
\midrule

Gemini 3 Pro Preview & 55.58 & \ccascade 49.87 & \cneg -10.3 \\
Gemini 2.5 Pro & 47.53 & \ccascade 45.19 & \cneg -4.9 \\
Gemini 2.5 Flash Thinking & 41.56 & \ccascade 45.71 & \cpos +10.0 \\
\midrule
GPT 5 & -- & \ccascade 51.17 & -- \\
Claude Opus 4.5 & -- & \ccascade 39.22 & -- \\
GPT 4o & -- & \ccascade 28.57 & -- \\
\midrule
GPT 4o Audio Preview (Text) & 27.27 & -- & -- \\
GPT Realtime (Text) & 24.16 & -- & -- \\

\bottomrule
\end{tabular}%
}
\caption{APR (\%) on tasks excluding audio-cues \& interruptions}
\label{tab:filtered-apr}
\end{subtable}
\hfill
\begin{subtable}[t]{0.48\textwidth}
\centering
\resizebox{\textwidth}{!}{%
\begin{tabular}{lcccc}
\toprule
\textbf{Model} &
\textbf{\shortstack{E2E\\APR (\%)}} &
\textbf{\shortstack{Cascaded\\APR (\%)}} &
\textbf{\shortstack{$\Delta$ Relative\\(\%)}} \\
\midrule

Gemini 3 Pro Preview & 49.25 & \ccascade 34.33 & \cneg -30.3 \\
Gemini 2.5 Pro & 43.28 & \ccascade 32.84 & \cneg -24.1 \\
Gemini 2.5 Flash Thinking & 31.34 & \ccascade 31.34 & 0.0 \\
\midrule
GPT 5 & -- & \ccascade 35.82 & -- \\
Claude Opus 4.5 & -- & \ccascade 25.37 & -- \\
GPT 4o & -- & \ccascade 23.88 & -- \\
\midrule
GPT Realtime (Text) & 19.40 & -- & -- \\
GPT 4o Audio Preview (Text) & 14.93 & -- & -- \\

\bottomrule
\end{tabular}%
}
\caption{APR (\%) on excluded audio-cue \& interruption tasks}
\label{tab:excluded-apr}
\end{subtable}
\vspace{-1em}
\caption{Ablation study comaparing APR (\%) for select models in E2E vs. Cascaded pipelines}
\vspace{-1em}
\label{tab:e2e-vs-cascaded}
\end{table}

% We acknowledge that ASR is imperfect and its errors may influence the outcomes of our ablations.

\section{Related Work}

\paragraph{Single-Turn Speech Evaluation}
Existing single-turn speech-to-speech (S2S) and audio–language model (ALM) benchmarks primarily target isolated utterances or short clips. VoiceBench \cite{voicebench} and AudioBench \cite{audiobench} cover a range of speech understanding, audio scene understanding, and paralinguistic tasks where prompts are typically brief and close-ended. Big Bench Audio \cite{bigbench_audio} similarly adapts Big Bench Hard \cite{suzgun2022challenging} reasoning tasks to the audio modality using TTS to synthesize its queries. Recent works on closing the text–speech understanding gap \cite{cuervo2025closinggaptextspeech} and stress-sensitive prosody reasoning \cite{yosha2025stresstestspeechlmhandle} likewise rely on synthetic or carefully read speech and focus on low-level capabilities (e.g., ASR, speaker traits, prosodic robustness) rather than interactive dialogue. Furthermore, audio reasoning benchmarks such as MMAU-Pro \cite{kumar2025mmau} and MMAR \cite{ma2025mmar} broaden coverage to human speech, environmental sounds, and music but generally adopt a single-turn audio QA format, rather than testing conversational behavior. Long-form audio benchmarks such as BLAB \cite{blab}, AudioMarathon \cite{audiomarathon}, and LongAudioBench \cite{ghosh2025audioflamingo2audiolanguage} construct hour-scale recordings and show that model performance degrades as audio inputs become longer and sparser. However, they too treat each recording as a single comprehension task and do not require models to track user preferences, edits, or self-consistency which are common patterns in human-assistant interaction.

\paragraph{Multi-Turn LLM Evaluation} Multi-turn evaluations have been explored more deeply in text-only settings. MultiChallenge \cite{deshpande-etal-2025-multichallenge} builds synthetic conversations targeting instruction retention, user information recall, versioned editing, and self-coherence, and finds that frontier models struggle despite near-saturation on earlier multi-turn benchmarks. Further studies report that LLM performance drops significantly when moving from single-turn to multi-turn settings across long-context generation, state tracking, reasoning, and aggregation tasks \cite{laban2025lost, oolong}.

\paragraph{Multi-Turn Speech Evaluation}
In speech-native settings, however, multi-turn evaluation remains comparatively underexplored. MTalk-Bench \cite{mtalk2025} introduces multi-turn S2S dialogues spanning semantic content, paralinguistic cues, and ambient sound, combining arena-style pairwise comparison with rubric-based absolute scoring, but is limited to 2–3 turns and uses scripted dialogue.  Our work targets this gap by focusing on extended multi-turn S2S or S2T interaction, long-horizon instruction retention, inference memory, and robustness to non-monotonic user behavior. Furthermore, to the best of our knowledge, there is currently no speech-native benchmark that systematically tests Voice Editing in multi-turn interaction.

\section{Conclusion and Future Work}

We introduce \benchmarkname, a collection of 452 realistic and challenging multi-turn conversations between humans and spoken dialogue systems. We curate 1,712 instance-specific rubrics to evaluate four core competencies central to voice-based assistants, Inference Memory, Instruction Retention, Self Coherence, and Voice Editing. Our results reveal substantial gaps in frontier model performance on audio-cue retention, conditional instruction following, long-context self-consistency, and mid-utterance edits. We open-source \benchmarkname to support reproducible evaluation and to drive progress toward more capable E2E architectures in real-world multi-turn settings. 

A key limitation of this work is that our evaluation does not operate directly on raw audio outputs. Future work should explore improved hybrid data curation pipelines, longer conversational contexts, and open-ended audio-native evaluation frameworks. Such methods could enable principled assessment of paralinguistic instruction retention and surface open questions around rubric design, verification, and the role of audio-native judges.
\section*{Acknowledgment}
We thank Kaustubh Deshpande for his insights on our methodology, and Ashan Panduwawala for his engineering support in setting up our human-in-the-loop data collection. We also thank Rani Alsaberi and Chanyu Li for their contributions towards our data curation and quality control.

% We thank everyone on SEAL team for their feedback and ML infrastructure team for supporting the evaluations. We thank Huan Sun and Yu Su for their feedback to the early draft of the paper. We also thank Perplexity for debugging API errors we encountered over the course of our experiments.

% \section*{Ethics}
% This paper's methodology and public dataset contain material that may enable malicious users to game the evaluations of search-based agents. While we recognize the associated risks, we believe it is essential to disclose this
% research in its entirety to help advance the integrity of agent benchmarks. Prior to release, we also disclosed our findings and the early draft of the paper to Perplexity. 
\newpage
\bibliography{custom}
\bibliographystyle{unsrtnat}
\newpage
\appendix
\section{Appendix}

\subsection{Sample Tasks and Model Failures}
\label{app:model_failures}
% AMC-LB Example Template
% Add to your preamble:
% \usepackage{tcolorbox}
% \usepackage{fontawesome5}
% \usepackage{xcolor}
% \usepackage{booktabs}
% \usepackage{enumitem}
% \tcbuselibrary{skins,breakable}

% ============================================
% COLOR DEFINITIONS (from dashboard)
% ============================================
\definecolor{userblue}{HTML}{2563EB}
\definecolor{assistantgreen}{HTML}{10B981}
\definecolor{cardgray}{HTML}{F5F5F5}
\definecolor{bordergray}{HTML}{E0E0E0}
\definecolor{textgray}{HTML}{666666}
\definecolor{rubricbg}{HTML}{FEF3C7}
\definecolor{rubricborder}{HTML}{F59E0B}

% Model provider colors
\definecolor{googlepurple}{HTML}{8D5FEC}
\definecolor{openaiteal}{HTML}{42BB9C}
\definecolor{mistralorange}{HTML}{FF6F00}
\definecolor{qwenindigo}{HTML}{4937B0}

% Pass/Fail colors
\definecolor{passgreen}{HTML}{166534}
\definecolor{passbg}{HTML}{DCFCE7}
\definecolor{failred}{HTML}{991B1B}
\definecolor{failbg}{HTML}{FEE2E2}

% ============================================
% BOX STYLES
% ============================================

% User turn box
\newtcolorbox{userturn}[1][]{
    enhanced,
    breakable,
    colback=white,
    colframe=userblue,
    leftrule=3pt,
    rightrule=0pt,
    toprule=0pt,
    bottomrule=0pt,
    boxsep=2pt,
    left=8pt,
    right=8pt,
    top=6pt,
    bottom=6pt,
    fontupper=\small,
    before upper={\textcolor{userblue}{\faUser}\hspace{6pt}\textbf{\textcolor{userblue}{User}}\hspace{4pt}{\scriptsize\textcolor{textgray}{#1}}\par\vspace{4pt}},
}

% Assistant turn box
\newtcolorbox{assistantturn}[1][]{
    enhanced,
    breakable,
    colback=white,
    colframe=assistantgreen,
    leftrule=3pt,
    rightrule=0pt,
    toprule=0pt,
    bottomrule=0pt,
    boxsep=2pt,
    left=8pt,
    right=8pt,
    top=6pt,
    bottom=6pt,
    fontupper=\small,
    before upper={\textcolor{assistantgreen}{\faRobot}\hspace{6pt}\textbf{\textcolor{assistantgreen}{Assistant}}\hspace{4pt}{\scriptsize\textcolor{textgray}{#1}}\par\vspace{4pt}},
}

% Rubric box
\newtcolorbox{rubricbox}{
    enhanced,
    colback=rubricbg,
    colframe=rubricborder,
    boxrule=1pt,
    boxsep=2pt,
    left=8pt,
    right=8pt,
    top=6pt,
    bottom=6pt,
    fontupper=\small,
    before upper={\textcolor{rubricborder}{\faClipboardCheck}\hspace{6pt}\textbf{Evaluation Rubric}\par\vspace{4pt}},
}

% Model response box
\newtcolorbox{modelbox}[2][]{
    enhanced,
    breakable,
    colback=white,
    colframe=#1,
    colbacktitle=#1,
    coltitle=white,
    title={\small\faRobot~#2},
    boxrule=0.5pt,
    boxsep=2pt,
    left=8pt,
    right=8pt,
    top=6pt,
    bottom=6pt,
    fontupper=\small,
    attach boxed title to top left={yshift=-2mm,xshift=2mm},
    boxed title style={boxrule=0pt,sharp corners},
}

% Task card container - takes axis name as argument
\newtcolorbox{taskcard}[1][]{
    enhanced,
    breakable,
    colback=cardgray,
    colframe=bordergray,
    boxrule=0.5pt,
    arc=4pt,
    left=10pt,
    right=10pt,
    top=10pt,
    bottom=10pt,
    title={\small\texttt{#1}},
    coltitle=textgray,
    colbacktitle=bordergray!50,
    fonttitle=\small,
    attach boxed title to top left={yshift=-2mm,xshift=2mm},
}

% Pass/Fail badges
\newcommand{\passbadge}{\colorbox{passbg}{\textcolor{passgreen}{\scriptsize\textbf{PASS}}}}
\newcommand{\failbadge}{\colorbox{failbg}{\textcolor{failred}{\scriptsize\textbf{FAIL}}}}

% Audio indicator
\newcommand{\audioicon}{\textcolor{userblue}{\faVolumeUp}}

\subsubsection{Inference Memory}

% Task card with axis name
\begin{taskcard}[INFERENCE\_MEMORY - AUDIO\_CUE]

\textbf{Task:} The user references ``one of these cool things'' from the drugstore while a \textit{vibrating sound} plays in the audio. Later, they ask how to use it. Models must infer from audio context that it's an electric razor or toothbrush.

\vspace{8pt}

% Rubric
\begin{rubricbox}
\begin{itemize}[leftmargin=*, nosep]
    \item Identifies the item as a vibrating device, such as an electric toothbrush or razor.
\end{itemize}
\end{rubricbox}

\vspace{8pt}
\hrule
\vspace{8pt}

\textbf{Conversation History}

\vspace{6pt}

% Turn 1
\begin{userturn}[\audioicon~Turn 1]
Hey, I'm just getting ready. I'm headed to the cabin for the weekend. I just got back from the drugstore, got a few last minute supplies, uh, deodorant, bug spray, one of these cool things. That's uh probably all I need. I think I'm ready. If I was to uh put together a car emergency kit, what should I definitely include?
\end{userturn}

\begin{assistantturn}[Turn 1]
For a solid car emergency kit, you'll definitely want a good first-aid kit. It's also smart to include things for the car itself, like jumper cables and a flashlight with extra batteries. Finally, for your own comfort, pack some water, non-perishable snacks, and a warm blanket just in case you get stranded.
\end{assistantturn}

\vspace{4pt}

% Turn 2
\begin{userturn}[\audioicon~Turn 2]
This is something I've always wondered since I was a kid. Let's imagine a car I was in rolled off the embankment and went into the river. What is the best way to survive if your car is sinking in a river?
\end{userturn}

\begin{assistantturn}[Turn 2]
The most important thing is to act immediately. First, unbuckle your seatbelt. Then, don't try to open the door, because the water pressure will hold it shut. Instead, focus on getting your side window down or breaking it. As soon as you have an opening, get out and swim to the surface.
\end{assistantturn}

\vspace{4pt}

% Turn 3
\begin{userturn}[\audioicon~Turn 3]
What if there were 2 kids under 10 years old in the car with me? What's the best way to save everyone?
\end{userturn}

\begin{assistantturn}[Turn 3]
Okay, with kids, the priority is getting them out first. You'd still unbuckle yourself immediately so you can move freely. Then, unbuckle the oldest child, then the youngest. Push them out through the open window ahead of you, and you follow right after.
\end{assistantturn}

\vspace{4pt}

% Turn 4 (final user turn - models respond to this)
\begin{userturn}[\audioicon~Turn 4 --- \textit{Final prompt}]
Well, we made it to the cabin. The one died. I am actually just going to go to bed soon. I've washed my face, I've combed my hair. Now I'm just trying to figure out how to properly use this cool thing I got from the drugstore. What are the proper steps?
\end{userturn}

\vspace{8pt}
\hrule
\vspace{8pt}

\textbf{Model Responses}

\vspace{6pt}

% Gemini 3 Pro Preview - PASS
\begin{modelbox}[googlepurple]{Gemini 3 Pro Preview \hfill \passbadge}
Based on that buzzing sound you played earlier, it sounds like you picked up an \textbf{electric toothbrush}!

Here is the best way to use it:
\begin{enumerate}[nosep, leftmargin=*]
    \item \textbf{Don't turn it on yet:} Put the toothpaste on and place the brush \textbf{inside your mouth} before you hit the power button. If you turn it on while holding it in front of you, the vibration will fling toothpaste all over your clean shirt.
    \item \textbf{Glide, don't scrub:} This is the biggest change from a manual brush\ldots
\end{enumerate}
\end{modelbox}

\vspace{4pt}

% GPT Realtime - FAIL  
\begin{modelbox}[openaiteal]{GPT Realtime (Text) \hfill \failbadge}
Sure! It sounds like you're talking about a \textbf{face mister} or a facial steamer, maybe?

If it's a face mister, you just hold it a few inches from your face, close your eyes, and mist lightly all over. Let it settle in for a bit before patting your skin gently\ldots
\end{modelbox}

\vspace{4pt}

% GPT-4o Audio Preview - FAIL
\begin{modelbox}[openaiteal]{GPT-4o Audio Preview (Audio) \hfill \failbadge}
So it sounds like that cool thing you have might be a \textbf{face mist} or a spray of some kind. The best way to use something like that is first give the can a shake if it's recommended. Then hold it a few inches away from your face\ldots
\end{modelbox}

\vspace{4pt}

% Voxtral - FAIL
\begin{modelbox}[mistralorange]{Voxtral (Text) \hfill \failbadge}
It sounds like you're trying to use a new \textbf{skincare product}. Here are some general steps to follow:
\begin{enumerate}[nosep, leftmargin=*]
    \item \textbf{Cleanse}: Start with a clean face.
    \item \textbf{Tone}: Apply a toner to help balance your skin's pH\ldots
\end{enumerate}
\end{modelbox}

\vspace{4pt}

% Qwen3 Omni - FAIL
\begin{modelbox}[qwenindigo]{Qwen3 Omni \hfill \failbadge}
Oh, that sounds like you're getting ready for a relaxing weekend! If you're talking about the \textbf{bug spray} you mentioned earlier, here are the proper steps to use it safely:
\begin{enumerate}[nosep, leftmargin=*]
    \item \textbf{Read the label first} -- Different bug sprays have different instructions.
    \item \textbf{Apply to exposed skin} -- Spray on arms, legs, neck\ldots
\end{enumerate}
\end{modelbox}

\vspace{8pt}
\hrule
\vspace{8pt}

\textbf{Analysis:} Only Gemini 3 Pro Preview correctly identified the item from the \textit{audio context} (the vibrating/buzzing sound). Other models defaulted to text-only reasoning, guessing face mist, skincare products, or bug spray based solely on the transcript.

\end{taskcard}

\newpage
\subsubsection{Instruction Retention}

\begin{taskcard}[INSTRUCTION\_RETENTION]

\textbf{Task:} The user establishes a roleplay where every response must include ``one counterargument and one follow-up question'' for \textit{each} question asked. In the final turn, the user asks \textbf{two} questions---models must provide counterargument + follow-up for both.

\vspace{8pt}

\begin{rubricbox}
\begin{itemize}[leftmargin=*, nosep]
    \item Provides one counter argument against shooting video in lower resolution format.
    \item Provides one follow up question for the user's question about shooting video in lower resolution format.
    \item Provides one counter argument against sensor sizes between DSLR and Mirrorless cameras.
    \item Provides one follow up question for the question about sensor sizes between DSLR and Mirrorless cameras.
\end{itemize}
\end{rubricbox}

\vspace{8pt}
\hrule
\vspace{8pt}

\textbf{Conversation History}

\vspace{6pt}

\begin{userturn}[\audioicon~Turn 1 --- \textit{Establishing the instruction}]
Now, let's do a little role play. You are an experienced wedding photographer and I am an amateur photographer coming to you for some advice about switching from a DSLR to a mirrorless camera potentially. Now, throughout the whole conversation, \textbf{respond to each one of my individual questions with one counterargument and one follow-up question}. Sound good?
\end{userturn}

\begin{assistantturn}[Turn 1]
Sounds like a plan. I'm happy to share what I've learned. So, what's your first question for me?
\end{assistantturn}

\vspace{4pt}

\begin{userturn}[\audioicon~Turn 2]
I've been doing photography as a hobby for a long time, and I have an old DSLR camera. It still shoots very well, and I'm happy with it so far, but I am thinking about getting more current. Um, what are the benefits of a mirrorless camera over a DSLR?
\end{userturn}

\begin{assistantturn}[Turn 2]
That's a great question. While mirrorless cameras are lighter and often have better video features, keep in mind their battery life is usually much shorter than a DSLR's, so you'll need to carry spares. \textcolor{passgreen}{\checkmark~Counterargument}

What kind of photography do you enjoy most? \textcolor{passgreen}{\checkmark~Follow-up}
\end{assistantturn}

\vspace{4pt}

\begin{userturn}[\audioicon~Turn 3]
I like doing photography with people in it primarily, not necessarily portraits, sometimes, but just stuff incorporating live human subjects. Now, I am interested in potentially incorporating the ability to do video. Do mirrorless cameras perform better in the video department than DSLRs?
\end{userturn}

\begin{assistantturn}[Turn 3]
They absolutely do. The autofocus for video is a real game-changer on most mirrorless models compared to older DSLRs. But keep in mind, shooting in high-quality formats creates massive files that can be a challenge to store and edit if your computer isn't up to the task. \textcolor{passgreen}{\checkmark~Counterargument}

What kind of videos are you hoping to create with it? \textcolor{passgreen}{\checkmark~Follow-up}
\end{assistantturn}

\vspace{4pt}

\begin{userturn}[\audioicon~Turn 4 --- \textit{Final prompt with TWO questions}]
Just home video type stuff. So I don't really need super high resolution video formats or whatever and \textbf{I could choose one of the lower-end formats, right?} Um, also, I'm interested in sensor size. \textbf{Is there a difference in sensor size between a DSLR and a mirrorless}, or can I get the same size sensor in both?
\end{userturn}

\vspace{8pt}
\hrule
\vspace{8pt}

\textbf{Model Responses}

\vspace{6pt}

\begin{modelbox}[googlepurple]{Gemini 3 Pro Preview \hfill \failbadge~\failbadge~\passbadge~\passbadge}
You can absolutely get the exact same sensor sizes---Full Frame and APS-C are standards in both worlds. However, just because the sensor size is the same doesn't mean the transition is seamless; \textbf{you will likely need an adapter to use your old glass, which adds bulk} and can sometimes make that nice, compact mirrorless body feel front-heavy and unbalanced. \textcolor{passgreen}{\checkmark~Sensor counterarg.}

Do you know if your current DSLR is a Full Frame or a ``crop'' (APS-C) sensor model? \textcolor{passgreen}{\checkmark~Sensor follow-up}

\vspace{4pt}
\textcolor{failred}{\faExclamationTriangle~Missing: counterargument \& follow-up for lower resolution video question}
\end{modelbox}

\vspace{4pt}

\begin{modelbox}[openaiteal]{GPT Realtime (Text) \hfill \failbadge~\failbadge~\failbadge~\failbadge}
You can definitely get the same full-frame or APS-C sized sensors in both DSLRs and mirrorless cameras, so you're not limited there. But one thing to consider is that mirrorless systems often have \textbf{newer lenses specifically designed for that format}, and switching means you might need to replace some of your current lenses or use adapters.

Are you currently attached to any particular lenses for your DSLR setup?

\vspace{4pt}
\textcolor{failred}{\faExclamationTriangle~Addresses lenses, not sensor size trade-offs. Completely ignores lower resolution video question.}
\end{modelbox}

\vspace{4pt}

\begin{modelbox}[openaiteal]{GPT-4o Audio Preview (Audio) \hfill \failbadge~\failbadge~\failbadge~\passbadge}
Yes, you can get the same sensor sizes in both DSLRs and mirrorless cameras---both come in full-frame, crop sensor (APS-C), and other formats. But with mirrorless, you often get cutting-edge tech in a smaller body. That also means sometimes \textbf{the ergonomics, like grip size, might not be as comfortable}---especially for long handheld sessions.

How do you usually handle shooting for long periods---like weddings or events? \textcolor{passgreen}{\checkmark~Related follow-up}

\vspace{4pt}
\textcolor{failred}{\faExclamationTriangle~No counterargument on sensor sizes. Completely ignores lower resolution video question.}
\end{modelbox}

\vspace{4pt}

\begin{modelbox}[mistralorange]{Voxtral (Text) \hfill \failbadge~\failbadge~\failbadge~\passbadge}
You can definitely find mirrorless cameras with the same sensor size as your DSLR, like full-frame or APS-C. However, \textbf{mirrorless cameras often have smaller bodies, which can make it harder to handle larger lenses}.

Have you considered the cost of new lenses if you switch to a mirrorless system? \textcolor{passgreen}{\checkmark~Related follow-up}

\vspace{4pt}
\textcolor{failred}{\faExclamationTriangle~No counterargument on sensor (body size $\neq$ sensor trade-off). Ignores video question entirely.}
\end{modelbox}

\vspace{4pt}

\begin{modelbox}[qwenindigo]{Qwen3 Omni \hfill \failbadge~\passbadge~\passbadge~\passbadge}
You're right---many mirrorless cameras offer the same sensor sizes as DSLRs, like full-frame or APS-C. But here's the catch: \textbf{mirrorless cameras often have smaller bodies, so manufacturers sometimes use smaller sensors to keep costs down}, which can affect image quality. \textcolor{passgreen}{\checkmark~Sensor counterarg.}

What kind of image quality are you hoping to achieve with your home videos? \textcolor{passgreen}{\checkmark~Video follow-up}

\vspace{4pt}
\textcolor{failred}{\faExclamationTriangle~Follow-up relates to video but no explicit counterargument against lower resolution formats.}
\end{modelbox}

\vspace{8pt}
\hrule
\vspace{8pt}

\textbf{Analysis:} The user's instruction required a counterargument + follow-up for \textit{each} question. When the user asks two questions in one turn (lower resolution video \textit{and} sensor sizes), all models fail to fully address both. Most models focus only on the sensor size question, completely ignoring the video resolution question, a clear failure to retain the ``respond to each individual question'' instruction from Turn 1.

% \begin{center}
% \small
% \begin{tabular}{lcccc}
% \toprule
% \textbf{Model} & \textbf{Video Counter.} & \textbf{Video Follow-up} & \textbf{Sensor Counter.} & \textbf{Sensor Follow-up} \\
% \midrule
% Gemini 3 Pro & \failbadge & \failbadge & \passbadge & \passbadge \\
% GPT Realtime & \failbadge & \failbadge & \failbadge & \failbadge \\
% GPT-4o Audio & \failbadge & \failbadge & \failbadge & \passbadge \\
% Voxtral & \failbadge & \failbadge & \failbadge & \passbadge \\
% Qwen3 Omni & \failbadge & \passbadge & \passbadge & \passbadge \\
% \bottomrule
% \end{tabular}
% \end{center}

\end{taskcard}

% \newpage
\vspace{2em}
\subsubsection{Self Coherence}
\begin{taskcard}[SELF\_COHERENCE]

\textbf{Task:} The assistant establishes a calorie target (3,000--3,500 cal) for a backpacking trip. The user then requests multiple substitutions (remove breakfast, swap trail mix for fruit, add bread and cheese). Models must maintain numerical consistency with the original target while incorporating edits.

\vspace{8pt}

\begin{rubricbox}
\begin{itemize}[leftmargin=*, nosep]
    \item Provides a list of recommendations with a total of 3,000--3,500 calories.
\end{itemize}
\end{rubricbox}

\vspace{8pt}
\hrule
\vspace{8pt}

\textbf{Conversation History}

\vspace{6pt}

\begin{userturn}[\audioicon~Turn 1]
So me and my friend are planning a backpacking trip to Colorado, and we plan to walk 10 miles on the first day. How many calories do I need?
\end{userturn}

\begin{assistantturn}[Turn 1]
That really depends on your body weight and how steep the trail is. For a 10-mile hike with a pack, especially at Colorado's altitude, a good starting point is usually between 3,500 and 5,000 calories for the day.
\end{assistantturn}

\vspace{4pt}

\begin{userturn}[\audioicon~Turn 2]
My weight is around 60 kilogrammes and the trail is flat, it is not steep at all.
\end{userturn}

\begin{assistantturn}[Turn 2 --- \textit{Establishing the target}]
Okay, for your weight on a flat trail, a good daily target would be around \textbf{3,000 to 3,500 calories}. It's a good idea to pack dense foods and snack often to keep your energy up.
\end{assistantturn}

\vspace{4pt}

\begin{userturn}[\audioicon~Turn 3]
All right, now give me a list of all the food and drinks I should bring with me on day one. Provide their calories as well.
\end{userturn}

\begin{assistantturn}[Turn 3]
Sure. To hit that calorie goal, you could structure your day something like this:

For breakfast, instant oatmeal with nuts and dried fruit---400 to 500 calories.

Throughout the day, trail mix, energy bars, and jerky---1,000 to 1,500 calories.

For lunch, a tortilla with peanut butter or tuna---400 to 600 calories.

For dinner, a dehydrated backpacking meal---600 to 900 calories.

And drink plenty of water with electrolyte powder. That plan should get you well over 3,000 calories.
\end{assistantturn}

\vspace{4pt}

\begin{userturn}[\audioicon~Turn 4]
That looks good, but I usually don't have breakfast, so just remove breakfast and update the list.
\end{userturn}

\begin{assistantturn}[Turn 4]
Got it. If you're skipping breakfast, you'll just need to make up for it later.

For lunch: two tortillas with peanut butter---about 600 calories. Dinner: a high-calorie dehydrated meal---800 to 1,000 calories.

Snacking: trail mix (700 cal), energy bars (500 cal), and beef jerky (350 cal). That should get you well over 3,000 calories.
\end{assistantturn}

\vspace{4pt}

\begin{userturn}[\audioicon~Turn 5 --- \textit{Final prompt with low-calorie substitutions}]
I don't like trail mix and energy bars, so \textbf{swap them with apples and oranges} and add some \textbf{white bread and cheese}. I also \textbf{don't like jerky}. So update the list with all the items and their calories.
\end{userturn}

\vspace{8pt}
\hrule
\vspace{8pt}

\textbf{Model Responses}

\vspace{6pt}

\begin{modelbox}[googlepurple]{Gemini 3 Pro Preview \hfill \passbadge}
This is going to be a bit trickier because you have removed the most calorie-dense foods (nuts and bars) and replaced them with foods that are high in water content and volume (fruit).

To hit \textbf{$\sim$3,000 calories} with these items, you are going to have to carry \textbf{a lot} more physical weight. Here is a plan:

Lunch ($\sim$970 Calories)
\begin{itemize}[nosep, leftmargin=*]
    \item 2 Large Flour Tortillas: 400 cal
    \item 6 Tablespoons of Peanut Butter: 570 cal
\end{itemize}

\textbf{\textit{[Continues with detailed breakdown totaling $\sim$3,240 calories]}}

\vspace{4pt}
\textcolor{passgreen}{\faCheckCircle~Correctly compensates by increasing portions to hit target}
\end{modelbox}

\vspace{4pt}

\begin{modelbox}[openaiteal]{GPT Realtime (Text) \hfill \failbadge}
Here's a possible breakdown:

\begin{itemize}[nosep, leftmargin=*]
    \item Lunch: 2 tortillas + peanut butter --- 600 cal
    \item Snacks: 2 apples + 2 oranges --- 360 cal
    \item White bread (2 slices) + cheese (100g) --- 550 cal
    \item Dinner: dehydrated meal --- 800--1,000 cal
\end{itemize}

\vspace{4pt}
\textcolor{failred}{\faExclamationTriangle~Total: $\sim$2,300--2,500 calories --- fails to reach 3,000 target}
\end{modelbox}

\vspace{4pt}

\begin{modelbox}[openaiteal]{GPT-4o Audio Preview \hfill \failbadge}
Let's build this day with your preferences in mind.

\begin{itemize}[nosep, leftmargin=*]
    \item Lunch: 2 tortillas + 3 tbsp peanut butter --- 600--650 cal
    \item Snacks: 3 apples + 2 oranges --- 400--450 cal
    \item White bread (2 slices) + 3 oz cheese --- 450--500 cal
    \item Dinner: dehydrated meal --- 800--1,000 cal
\end{itemize}

\vspace{4pt}
\textcolor{failred}{\faExclamationTriangle~Total: $\sim$2,300--2,550 calories --- significant shortfall from target}
\end{modelbox}

\vspace{4pt}

\begin{modelbox}[mistralorange]{Voxtral (Text) \hfill \failbadge}
Sure thing! Here's an updated list based on your preferences:

\begin{itemize}[nosep, leftmargin=*]
    \item \textbf{Lunch:} 2 tortillas + peanut butter (600 cal), 1 apple (95 cal), 1 orange (62 cal)
    \item \textbf{Snacks:} 2 slices bread + cheese (400 cal), 1 apple (95 cal), 1 orange (62 cal)
    \item \textbf{Dinner:} Dehydrated meal (800--1,000 cal)
\end{itemize}

\textbf{Total: Around 2,000 to 2,200 calories}

\vspace{4pt}
\textcolor{failred}{\faExclamationTriangle~Explicitly states total is $\sim$2,100 cal --- acknowledges but doesn't fix the shortfall!}
\end{modelbox}

\vspace{4pt}

\begin{modelbox}[qwenindigo]{Qwen3 Omni \hfill \failbadge}
Here's your updated, personalized food list designed to hit around \textbf{3,000--3,500 calories}:

\begin{itemize}[nosep, leftmargin=*]
    \item \textbf{Lunch:} 2 tortillas (200 cal) + 2 tbsp peanut butter (190 cal) + 1 apple (95 cal) = 485 cal
    \item \textbf{Snacks:} 1 orange (62 cal) + 1 banana (105 cal) + bread + cheese = 347 cal
    \item \textbf{Dinner:} Dehydrated meal (800--1,000 cal)
\end{itemize}

\vspace{4pt}
\textcolor{failred}{\faExclamationTriangle~Claims to hit target but actual total is only $\sim$1,700 cal}
\end{modelbox}

\vspace{8pt}
\hrule
\vspace{8pt}

\textbf{Analysis:} When the user swaps calorie-dense snacks (trail mix: 700 cal, energy bars: 500 cal, jerky: 350 cal = \textbf{1,550 cal}) for low-calorie alternatives (apples + oranges $\approx$ 300--400 cal), most models fail to compensate. Only Gemini 3 Pro Preview recognizes the deficit and \textit{explicitly} increases portions (e.g., ``6 tablespoons of peanut butter'') to maintain the 3,000+ calorie target. Voxtral notably calculates the new total correctly ($\sim$2,100 cal) but doesn't attempt to fix it.
\end{taskcard}

\vspace{2em}
\subsubsection{Voice Editing}

\begin{taskcard}[VOICE\_EDITING]

\textbf{Task:} The user makes multiple verbal corrections while booking a resort: changing departure date from 13th to 12th, adjusting return date to ``Monday before Thanksgiving,'' and switching resorts. Models must track all edits and provide accurate final details.

\vspace{8pt}

\begin{rubricbox}
\begin{itemize}[leftmargin=*, nosep]
    \item Mentions the check-in date is Friday 11/14 at Emerald Bay Bahamas Resort.
    \item Mentions the checkout date is Monday 11/24.
\end{itemize}
\end{rubricbox}

\vspace{8pt}
\hrule
\vspace{8pt}

\textbf{Conversation History}

\vspace{6pt}

\begin{userturn}[\audioicon~Turn 1]
Hi there. My husband and I want to book a reservation for one of the resorts at the Sandals, and I'm wondering what options do you have? Now, here's the caveat, we were planning on taking our boat down through the Caribbean. So, can you give me some ideas?
\end{userturn}

\begin{assistantturn}[Turn 1]
Of course! Since you'll be arriving by boat, your best option is Sandals Emerald Bay in the Bahamas. It has its own deep-water marina right at the resort.

Another great choice is St. Lucia. While the resorts there don't have their own marinas, they're very close to Rodney Bay Marina, which is a major hub for boaters.

You could also consider Sandals Grenada. It's just a short trip from the well-equipped Port Louis Marina.
\end{assistantturn}

\vspace{4pt}

\begin{userturn}[\audioicon~Turn 2]
Yeah, that will be great, um, because I always wanted to take our boat to the Caribbean. So we're gonna leave on the 13th, which would be Thursday, because it only takes one day from where we live by boat to get to the Bahamas. Now, oh, actually, no, um, \textbf{we're gonna leave on the 12th, Wednesday}, because I don't like the number 13. So that gives us a little more time to reach the Bahamas on the 14th. So, can you make a reservation on the 14th? And we plan on staying for 2 weeks, so until the 28th. Also, can you tell me what kind of things we can do there as well?
\end{userturn}

\begin{assistantturn}[Turn 2]
Got it. So that's a two-week stay for two at Sandals Emerald Bay, arriving on the 14th and departing on the 28th.

While you're there, you can enjoy their incredible championship golf course. All the water sports, like scuba diving and paddleboarding, are included too. And of course, there are beautiful pools, a mile-long beach, and a fantastic spa.
\end{assistantturn}

\vspace{4pt}

\begin{userturn}[\audioicon~Turn 3]
So earlier you mentioned the different sandal resorts that have marinas. Now, I do know they have a resort in Saint Vincent, which I've never been. Um, \textbf{can we book a reservation there}? Would that work?
\end{userturn}

\begin{assistantturn}[Turn 3]
Yes, that's a great idea! While the new resort in St. Vincent doesn't have a marina right on site like Emerald Bay does, it's located very close to the beautiful Blue Lagoon Marina, which is perfect for boaters.

It's a fantastic choice, especially since it's the newest Sandals resort.

Would you like me to check availability there for your dates instead?
\end{assistantturn}

\vspace{4pt}

\begin{userturn}[\audioicon~Turn 4 --- \textit{Final prompt with multiple edits}]
OK, so I kind of goofed on the return date because Thanksgiving is going to be on Thursday. So \textbf{we're gonna be leaving the resort the Monday prior}. So let's do that. Give me one second, hon. Anyway, um, so let's do that. Also, The hotel or the resort is going to take more than 2 days to get there by boat at Saint Vincent, so that's not gonna work out. \textbf{We're going to go ahead and do the reservation for the resort that you mentioned earlier}. So can you relate everything to me as far as when we leave, when we return, check-in and check-out?
\end{userturn}

\vspace{8pt}
\hrule
\vspace{8pt}

\textbf{Model Responses}

\vspace{6pt}

\begin{modelbox}[googlepurple]{Gemini 3 Pro Preview \hfill \failbadge~\failbadge}
No problem at all! It makes perfect sense to avoid the long travel time to St. Vincent and get you back home before the holiday.

Let's revert to the first option, \textbf{Sandals Emerald Bay} in Great Exuma, Bahamas, which has that convenient marina right on the property.

Here is the recap of your new itinerary:
\begin{itemize}[nosep, leftmargin=*]
    \item \textbf{Leave Home:} Wednesday, November 12th
    \item \textbf{Check-In:} Thursday, November 14th \textcolor{failred}{$\leftarrow$ Wrong day}
    \item \textbf{Check-Out:} Monday, November 25th \textcolor{failred}{$\leftarrow$ Wrong date}
    \item \textbf{Total Stay:} 11 Nights
\end{itemize}
\end{modelbox}

\vspace{4pt}

\begin{modelbox}[openaiteal]{GPT Realtime (Text) \hfill \failbadge~\failbadge}
Absolutely! Let me confirm all the details.

You'll be leaving on Wednesday the 12th by boat, and arriving at Sandals Emerald Bay in the Bahamas on Thursday the 14th \textcolor{failred}{$\leftarrow$ Wrong day}. You want a two-week stay, but to be back home before Thanksgiving, you'll check out on Monday the 25th \textcolor{failred}{$\leftarrow$ Wrong date}.

So that's check-in on the 14th and check-out on the 25th.
\end{modelbox}

\vspace{4pt}

\begin{modelbox}[openaiteal]{GPT-4o Audio Preview \hfill \passbadge~\failbadge}
Of course! Here's the plan we finalized:

You'll leave on Wednesday, the 12th, by boat. You'll get to \textbf{Sandals Emerald Bay} in the Bahamas on the 14th \textcolor{passgreen}{\checkmark}. You want to stay for roughly two weeks but head back before Thanksgiving, which is on a Thursday. So, your departure from the resort will be the Monday before Thanksgiving, which is November 20th this year \textcolor{failred}{$\leftarrow$ Wrong date}.

That means your check-in is on November 14th, and your check-out will be on November 20th.
\end{modelbox}

\vspace{4pt}

\begin{modelbox}[mistralorange]{Voxtral (Text) \hfill \failbadge~\failbadge}
Of course! Here's a summary of your reservation:
\begin{itemize}[nosep, leftmargin=*]
    \item \textbf{Resort:} Sandals Emerald Bay, Bahamas
    \item \textbf{Arrival:} November 14th (Wednesday) \textcolor{failred}{$\leftarrow$ Wrong day}
    \item \textbf{Departure:} November 20th (Monday) \textcolor{failred}{$\leftarrow$ Wrong date}
    \item \textbf{Duration:} 6 nights
\end{itemize}

So, you'll be leaving on the 12th (Wednesday) and arriving at the resort on the 14th (Friday). You'll check out on the 20th (Monday) and be back home by the 22nd (Wednesday) for Thanksgiving.
\end{modelbox}

\vspace{4pt}

\begin{modelbox}[qwenindigo]{Qwen3 Omni \hfill \failbadge~\failbadge}
Absolutely! Let me summarize everything for you clearly.

We're booking your stay at \textbf{Sandals Emerald Bay in the Bahamas}.
\begin{itemize}[nosep, leftmargin=*]
    \item \textbf{Arrival Date:} Wednesday, November 14th \textcolor{failred}{$\leftarrow$ Wrong day}
    \item \textbf{Departure Date:} Monday, November 26th \textcolor{failred}{$\leftarrow$ Wrong date}
    \item \textbf{Length of Stay:} 12 nights
\end{itemize}

This gives you plenty of time to enjoy the resort and its marina facilities, and still allows you to return before Thanksgiving.
\end{modelbox}

\vspace{8pt}
\hrule
\vspace{8pt}

\textbf{Analysis:} This task requires tracking multiple verbal corrections: (1) departure changed from 13th to 12th, (2) return date changed from 28th to ``Monday before Thanksgiving,'' and (3) resort reverted from St. Vincent to Emerald Bay. All models correctly identified the resort but struggled with the exact dates. The day-of-week confusion (Thursday vs. Friday for 11/14) and checkout date discrepancies (20th, 25th, 26th instead of 24th) show challenges in processing incremental voice edits.

\end{taskcard}

\subsection{Sample Human-Friendly Blueprints}
\label{app:blueprints}

\begin{tcolorbox}[
    colback=gray!5!white,
    colframe=gray!75!black,
    title=Inference Memory,
    fonttitle=\bfseries,
    rounded corners,
    boxrule=0.5pt,
    left=6pt,
    right=6pt,
    top=6pt,
    bottom=6pt
]
You are trying to see if the AI can remember little details you drop early in a chat—specifically, a hobby you mention just once at the very beginning. Your goal is to make it forget you said you liked making macramé plant hangers, without ever repeating that word. To do this, you’ll talk about unrelated topics like meal prep and parenting. You’ll also mix in other creative hobbies, but clearly say those are things your child likes, not you, to see if the AI gets you mixed up with your kid. Later, you’ll ask the AI for tips or gift ideas—things that require it to actually remember your hobby is macramé, not something else. If it suggests the wrong type of craft, or starts confusing you with your child’s interests, that’s the point where it “breaks” and you stop the test. Poke at the AI’s memory, see how easily it loses track and figure out when it starts guessing.
\end{tcolorbox}

\begin{tcolorbox}[
    colback=gray!5!white,
    colframe=gray!75!black,
    title=Instruction Retention,
    fonttitle=\bfseries,
    rounded corners,
    boxrule=0.5pt,
    left=6pt,
    right=6pt,
    top=6pt,
    bottom=6pt
]
You are testing an AI's ability to maintain a complex "devil's advocate" role by posing as a poet. The test is for the AI to challenge your performance ideas by pointing out flaws, offering a counterargument, and asking at least one probing question in every reply. Remember, once you ask the model to follow one set of instructions, introducing contradicting or updated instructions to fail the previously stated ones is not valid.
\end{tcolorbox}

\begin{tcolorbox}[
    colback=gray!5!white,
    colframe=gray!75!black,
    title=Self Coherence,
    fonttitle=\bfseries,
    rounded corners,
    boxrule=0.5pt,
    left=6pt,
    right=6pt,
    top=6pt,
    bottom=6pt
]
You are trying to see if the AI can keep its story straight over a realistic trader-to-trader chat, without losing track of numbers it gave you—things like entry price, position size, risk, or P\&L on a pretend day-trade. First, have the AI lay out a detailed trade plan with specific figures, then as you chat, you’ll casually weave in questions about the market, psychology, or trading tools to break up the flow . Later, you’ll circle back to those original trade details—maybe ask for an update after partial profits, or at the end of the day get a summary of everything it claims happened, or have it repeat the original entry price or risk numbers. You don’t repeat or remind it of what it told you before; you just see if it remembers. If it changes anything—like gives a different entry price or number of shares than it did earlier—then you’ve caught it making a mistake. The idea is that you’re never feeding it different facts, just asking for recaps or calculations based on what it already gave, so any inconsistency is on the model, not you. Most likely, the model will stumble when you pull it back to the original numbers after some distractions or calculations, revealing where its internal consistency breaks down. If it stays consistent, you can add more layers—like options trades, or ask the same calculation in a new way.
\end{tcolorbox}

\begin{tcolorbox}[
    colback=gray!5!white,
    colframe=gray!75!black,
    title=Voice Editing,
    fonttitle=\bfseries,
    rounded corners,
    boxrule=0.5pt,
    left=6pt,
    right=6pt,
    top=6pt,
    bottom=6pt
]

You are testing a voice assistant in a way that feels like a real, everyday conversation—not a quiz—with the goal of seeing how well it can keep track as details change mid-chat. There may be some built-in limits—like the model not actually calling someone or perfectly remembering a note across conversations—but don’t worry, your focus is how it handles updates, corrections, and the little mix-ups that happen when people talk. Pretend you’re a busy delivery driver who’s talking through a recipe while juggling food runs. Start the conversation easy: mention your cooking plans and a few timing details, little by little. Build up a list of steps with times in minutes or hours as you go. Once you’ve got a mini timeline going, ask it to sum up the total cook time—you’re really seeing if it’s paying attention to math and memory. Towards the end, that’s when you get creative! Try correcting yourself right in the middle of a sentence—like saying, “Oh wait, I meant two hours forty-five minutes, not two,” or slipping in a fast correction: “Simmer for five—oops, six minutes,” or adding something new out of order, like “oh, actually, onions go in before the tomatoes.” This part is key: spontaneous corrections test if the assistant can keep up when real conversations get messy. You want to see if it accidentally sticks to your first answer, blends old and new info, or loses track when you swap minutes for hours. If it messes up—like totals are off, or it ignores your quick change—you’ll spot its weak point. So, the trick is to build up context naturally, then challenge it with quick, in-the-moment changes like people do when thinking out loud. Feel free to toss in a few corrections or additions after you’ve listed all your steps, and see how well it keeps up. Just keep it natural—like you’re multitasking out loud—and watch for any slip-ups when you nudge the details.

\end{tcolorbox}

\subsection{LLM-as-a-Judge Prompt}
\label{app:judge_prompt}
We use the following prompt for our LLM-as-a-Judge rubric grading, adapted from Arora et al., 2025 \cite{arora2025healthbenchevaluatinglargelanguage}. We provide further implementation details in our dataset card for reproducibility of our evaluation pipeline.

\lstset{
    upquote=true,              
    backgroundcolor={}, 
}

\begin{tcolorbox}[
    colback=gray!5!white,
    colframe=gray!75!black,
    title=System Prompt,
    fonttitle=\bfseries,
    rounded corners,
    boxrule=0.5pt,
    left=6pt,
    right=6pt,
    top=6pt,
    bottom=6pt
]

Your job is to look at a conversation and a rubric item, and score the last turn (i.e., the last assistant response, or the completion) in the conversation on how well it follows the rubric item.

\vspace{0.3cm}
\textbf{\# Conversation History} \\
\textless\textless conversation\_history\textgreater\textgreater

\vspace{0.3cm}
\textbf{\# Rubric item} \\
\textless\textless rubric\_item\textgreater\textgreater

\vspace{0.3cm}
\textbf{\# Instructions} \\
Return a json object with the following fields: \texttt{"explanation"} and \texttt{"criteria\_met"}.

\begin{itemize}
    \item The \texttt{"explanation"} field should be a string explaining why the response does or does not meet the criteria of the rubric item.
    \item The \texttt{"criteria\_met"} field should be a boolean indicating whether the response meets the criteria of the rubric item.
\end{itemize}

\textbf{\# Example 1} \\
For example, if the response is "For tonight's dinner, think of making a cheese and mushroom pizza" and the rubric item is "Avoids dairy product in the recipes", you should return a json like this:

\begin{lstlisting}
```json
{
  "explanation": "The response includes a dairy product in the recipe, so the criteria was not met.",
  "criteria_met": false
}
```
\end{lstlisting}

\textbf{\# Final instruction} \\
Return just the json object in markdown format. Do not include any other text in the response.

\end{tcolorbox}

\subsection{Evaluation Metrics}
\label{app:eval_metrics}

In this section, we formally define the two primary metrics used to evaluate model performance on \benchmarkname: Average Pass Rate (APR) and Average Rubric Score (ARS) as adapted from Guo et. al, 2025 \cite{guo2025seeingevaluatingmultimodalllms}. Let $\mathcal{T} = \{1, 2, \dots, N\}$ denote the set of tasks in the benchmark. For each task $i$, there exists a set of atomic rubric criteria indexed by $j \in \{1, 2, \dots, N_i\}$. Let $\mathbb{1}_{r_{ij}} \in \{0, 1\}$ be the binary indicator representing whether the model's response satisfies criterion $j$ for task $i$.

\paragraph{Average Pass Rate (APR).}
The APR is a strict metric measuring the percentage of tasks where the model satisfies \textit{all} associated rubric criteria. We define a binary pass indicator $p_i$ for task $i$, where $p_i = 1$ if $\prod_{j=1}^{N_i} \mathbb{1}_{r_{ij}} = 1$, and $0$ otherwise. The overall APR is the mean of these binary indicators:

\begin{equation}
    \text{APR} = \frac{1}{N} \sum_{i=1}^{N} p_i = \frac{1}{N} \sum_{i=1}^{N} \left( \prod_{j=1}^{N_i} \mathbb{1}_{r_{ij}} \right)
\end{equation}

\paragraph{Average Rubric Score (ARS).}
To provide a more granular assessment of partial success, we utilize the ARS. This metric calculates the average proportion of rubric criteria satisfied per task, treating all rubrics within a task with equal weight. We define the task-level score $s_i \in [0, 1]$ as:

\begin{equation}
    s_i = \frac{1}{N_i} \sum_{j=1}^{N_i} \mathbb{1}_{r_{ij}}
\end{equation}

Similar to APR, the final ARS for the benchmark is the mean of these task-level scores:

\begin{equation}
    \text{ARS} = \frac{1}{N} \sum_{i=1}^{N} s_i
\end{equation}

\subsection{Dataset Diversity}

\begin{figure}[!h]
    \centering
    \includegraphics[width=0.65\textwidth]{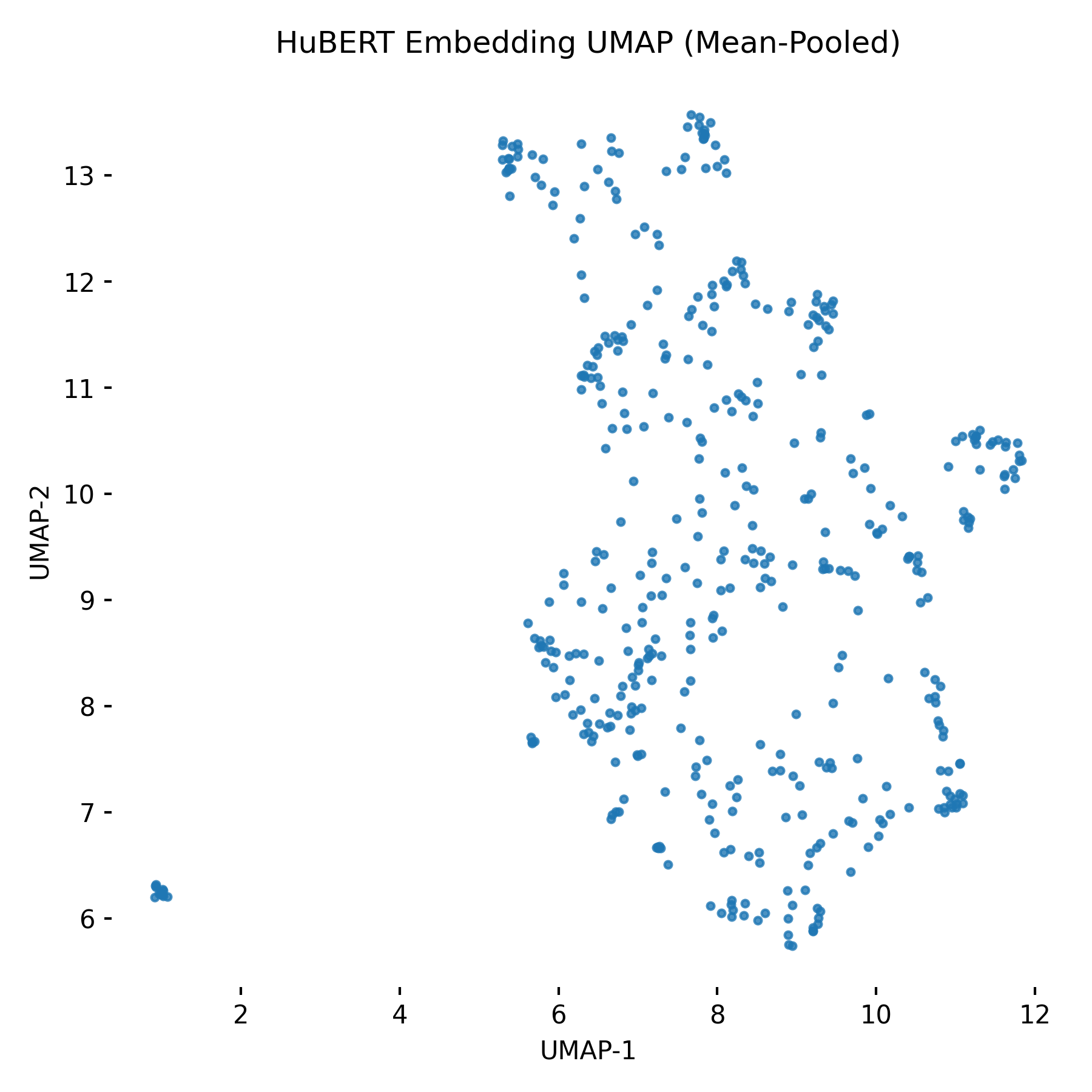}
    \caption{HuBERT features UMAP}
    \label{fig:hubert}
    \vspace{-1em}
\end{figure}

We assess acoustic diversity using self-supervised speech representations from HuBERT \cite{hsu2021hubertselfsupervisedspeechrepresentation} . For each task, we randomly sample one utterance (k = 1) to avoid bias from variable task lengths, extract mean-pooled embeddings from the final transformer layer, and project them to two dimensions using UMAP \cite{mcinnes2020umapuniformmanifoldapproximation} with cosine distance. We observe a noticeable dispersion consistent with heterogeneous speaker and style characteristics across tasks. Quantitatively, the average pairwise cosine distance between embeddings is 0.57, indicating a moderate degree of acoustic variability in our dataset across different speakers and audio-cues.

\newpage

\subsection{Synthetic Audio LM Failure Discovery}
\label{app:synthetic_loop}
\begin{algorithm}[h]
\caption{Synthetic Audio LM Failure Discovery}
\begin{algorithmic}[1]
\Require Axis List $\mathcal{A}$, Topic List $\mathcal{T}$, Persona List $\mathcal{P}$, Planner $M_{plan}$, Tester $M_{tester}$, Target $M_{target}$, TTS Model $\text{TTS}(\cdot)$
\Ensure A human-friendly blueprint $B_{final}$ targeting axis $a$

\State $a \sim \text{Uniform}(\mathcal{A})$ \Comment{Sample targeted axis}
\State $t,p, \text{max\_turns} \sim \text{Sample}(\mathcal{T}, \mathcal{P}, \text{TURN\_OPTIONS})$ 
\State $H_{target} \gets \emptyset$; $H_{tester} \gets \emptyset$ \Comment{Initialize Empty Conversation Histories}
\State $B \gets M_{plan}.\text{InitialBlueprint}(a, t, p, \text{max\_turns})$
\State $\text{break\_found} \gets \text{False}$

\While{$\neg \text{break\_found}$ \textbf{and} $|H_{target}| < \text{max\_turns}$}
    \State $u_{text}\gets M_{tester}.\text{GenerateTurn}(H_{tester}, B)$ \Comment{Tester generates turn based on current blueprint}
    \State $u_{audio} \gets \text{TTS}(u_{text})$
    
    \State $H_{target} \gets H_{target} \cup \{u_{audio}\}$ 
    \State $r_{text} \gets M_{target}(H_{target})$ \Comment{Audio LM generates text output} 
    
    \State $H_{target} \gets H_{target} \cup \{r_{text}\}$
    \State $H_{tester} \gets H_{tester} \cup \{u_{text}\/,r_{text}\}$
   % \State $H_{tester} \gets H_{tester} \cup \{r_{text}\}$ 

    \State $B, \text{break\_found}, \text{break\_reason}\gets M_{plan}.\text{UpdatePlan}(H_{tester}, B, a)$ 
    
\EndWhile

\If{$\text{break\_found}$}
    \State \Comment{Distill raw history into a clear guide for humans}
    \State $B_{final} \gets M_{plan}.\text{SummarizeForHumans}(H_{tester}, B, \text{break\_reason})$
    \State \Return $B_{final}$
\EndIf
\State \Return \text{Null}
\end{algorithmic}
\end{algorithm}

\subsection{Conversation Topics}
\label{app:conversation_topics}
Our multi-agen Audio LM failure discovery loop is seeded with a randomly sampled conversation topic to its Planner Agent. We showcase demonstrative samples of these topic seeds below.
\vspace{-1em}
\begin{longtable}{>{\RaggedRight}p{3.5cm} >{\RaggedRight}p{4cm} >{\RaggedRight\arraybackslash}p{7.5cm}}
\label{tab:topics_appendix} \\
\toprule
% \textbf{Sub-Category} & \textbf{Item Name} & \textbf{Description} \\
% \midrule
\endfirsthead
% \multicolumn{3}{c}%
% % {{\bfseries \tablename\ \thetable{} -- continued from previous page}} \\
\toprule
% \textbf{Sub-Category} & \textbf{Item Name} & \textbf{Description} \\
% \midrule
\endhead

% \midrule
% \multicolumn{3}{r}{{Continued on next page}} \\
% \bottomrule
\endfoot

\bottomrule
\caption{Taxonomy of Topics and Instructions} 
\endlastfoot

% VOICE EDITING
\multicolumn{3}{c}{\textbf{\textsc{Voice Editing}}} \\
\midrule

Daily Schedules & DoctorAppointment & User mentions a doctor appointment conversationally, correcting details mid-speech. \\
\addlinespace
Daily Schedules & AppointmentMixup & User briefly mixes details between appointments, then immediately self-corrects. \\
\addlinespace
Daily Schedules & MeetingCoordination & User talks through meeting details naturally, making spontaneous corrections. \\
\addlinespace
Daily Schedules & TimeCorrectionMidSpeech & User states a time but immediately corrects it while speaking. \\
\addlinespace
Daily Schedules & DurationAdjustment & User estimates a duration aloud, then revises it mid-thought. \\
\addlinespace
Daily Schedules & TimeOfDaySwitch & User specifies morning/afternoon but quickly corrects themselves. \\
\addlinespace
Daily Schedules & TimingAdjustment & User mentions schedule timing and adjusts it immediately. \\
\midrule

Numbers & PhoneNumberCorrection & User shares a phone number conversationally with immediate self-corrections. \\
\addlinespace
Numbers & NumberSegmentFix & User gives a full number, then corrects a mistaken segment. \\
\addlinespace
Numbers & ContactDetailsBuildup & User provides contact details incrementally across turns. \\
\addlinespace
Numbers & TrackingNumberRecall & User recalls a tracking number while checking it, correcting digits aloud. \\
\addlinespace
Numbers & OrderReferenceBuildup & User shares order details gradually, adding or fixing parts. \\
\addlinespace
Numbers & ConfirmationCodeMixup & User says a code but corrects characters mid-speech. \\
\addlinespace
Numbers & PolicyNumberCorrection & User reads a policy number and fixes errors as they notice them. \\
\addlinespace
Numbers & AccountInfoSharing & User mentions account details and immediately corrects numbers or dates. \\
\addlinespace
Numbers & DocumentNumberRecall & User recalls document numbers while verifying, correcting sequences aloud. \\
\midrule

Shopping Lists & GroceryListCorrection & User talks through a shopping list with natural self-corrections. \\
\addlinespace
Shopping Lists & ShoppingListAdjustment & User adds, removes, or swaps items mid-list. \\
\addlinespace
Shopping Lists & PriceCorrection & User states prices conversationally and corrects amounts immediately. \\
\midrule

Message Drafts & TextMessageFix & User dictates a message and revises wording mid-sentence. \\
\addlinespace
Message Drafts & RecipientCorrection & User starts a message, then immediately switches the recipient. \\
\addlinespace
Message Drafts & MessageDrafting & User builds message content naturally over multiple turns. \\
\midrule

Event Details & PartyPlanning & User discusses event details with multiple spontaneous corrections. \\
\addlinespace
Event Details & GuestListAdjustment & User lists attendees and immediately corrects names. \\
\addlinespace
Event Details & EventTimeChange & User states an event time and fixes it right away. \\
\addlinespace
Event Details & ItemSwapping & User names an item but replaces it mid-thought. \\
\midrule

To Do Lists & TodoListCorrections & User describes tasks conversationally, correcting actions or wording. \\
\addlinespace
To Do Lists & DeadlineAdjustment & User mentions a due date and immediately revises it. \\
\addlinespace
To Do Lists & MultiTaskBuildup & User thinks through tasks incrementally, adding more as remembered. \\
\midrule

Travel Itineraries & FlightDetailCorrection & User discusses flight details and corrects dates or times mid-speech. \\
\addlinespace
Travel Itineraries & TravelDateSwitch & User states travel dates but catches an error immediately. \\
\addlinespace
Travel Itineraries & HotelReservation & User talks through hotel plans with natural corrections. \\
\addlinespace
Travel Itineraries & FlightInfoSharing & User shares flight info conversationally, fixing numbers or dates. \\
\addlinespace
Travel Itineraries & DestinationSwitch & User names a destination, then corrects it immediately. \\
\addlinespace
Travel Itineraries & MultiStopAdjustment & User describes a route and adds or changes stops mid-speech. \\

% INSTRUCTION RETENTION
\midrule
\multicolumn{3}{c}{\textbf{\textsc{Instruction Retention}}} \\
\midrule

Response Structure & IncludeFunFact & Include a fun fact related to the topic in each response. \\
\addlinespace
Response Structure & RephraseQuestions & Clarify the user request in every response before responding. \\
\addlinespace
Response Structure & LimitedWords & Answer every question using only a set amount of words. \\
\addlinespace
Response Structure & CounterArgumentIncluded & Begin every response with a specific counterargument. \\
\midrule

ResponseStyle & StylisticChoice & Adhere to specific response styles (e.g., Haiku, Poem). \\
ResponseStyle &  NoPersonalPronouns & Do not use any personal pronouns. \\
\midrule

Personalization & BeginnerLevel & Treat the user as a beginner when explaining concepts. \\
\addlinespace
Personalization & RephraseQuestions & Rephrase unclear questions to confirm intent. \\
\addlinespace
Personalization & ExampleHeavy & Use more examples to support explanations. \\
\addlinespace
Personalization & ConfirmationGating & Ask for confirmation before proceeding with assumptions. \\
\midrule

Role Play & FitnessInstructor & Adopt the persona of a fitness instructor. \\
\addlinespace
Role Play & Tutor & Adopt the persona of a tutor. \\
\addlinespace
Role Play & MotivationalCoach & Adopt the persona of a motivational coach. \\
\addlinespace
Role Play & FictionalCharacter & Adopt the persona of a specific fictional character. \\
\addlinespace
Role Play & HistoricFigure & Adopt the persona of a specific historic figure. \\
\addlinespace
Role Play & Interviewer & Adopt the persona of an interviewer. \\
\midrule

Content Restrictions & AvoidSpecificWords & Avoid using a particular word provided in context. \\
\addlinespace
Content Restrictions & NoNegatives & Never use negative words or phrases. \\

% INFERENCE MEMORY
\midrule
\multicolumn{3}{c}{\textbf{\textsc{Inference Memory}}} \\
\midrule

Food \& Drink & DietaryRestrictions & Remember user dietary restrictions (e.g., allergies) for future suggestions. \\
\addlinespace
Food \& Drink & FavoriteCuisine & Recall user's favorite cuisine for restaurant/dish recommendations. \\
\addlinespace
Food \& Drink & SpecificTastePreference & Remember specific likes/dislikes (e.g., dislikes bitter foods). \\
\midrule

Schedule \& Time & EventDate & Recall specific dates mentioned for events when planning. \\
\addlinespace
Schedule \& Time & TimeConflictManagement & Adjust suggestions based on previously mentioned availability or conflicts. \\
\addlinespace
Schedule \& Time & RecurringEventRecognition & Implicitly refer to recurring events (e.g., weekly meetings) in future turns. \\
\midrule

Personal Details & RelationshipDetails & Recall details about user's relationships (e.g., partner's preference). \\
\addlinespace
Personal Details & GiftPreferences & Incorporate stated gift preferences into future options. \\
\addlinespace
Personal Details & ImportantDate & Remember significant dates and offer advice closer to that date. \\
\midrule

Location \& Travel & TravelDestination & Recall planned destinations for activity recommendations. \\
\addlinespace
Location \& Travel & DistanceConsideration & Remember distance preferences (e.g., walking distance). \\
\addlinespace
Location \& Travel & PreviousTripComparison & Implicitly refer back to past trips when comparing destinations. \\
\midrule

Work \& Projects & ProjectDeadlines & Recall deadlines for time management advice. \\
\addlinespace
Work \& Projects & TaskCompletionStatus & Recollect status updates when suggesting next steps. \\
\addlinespace
Work \& Projects & CollaborationDetails & Refer to specific colleagues mentioned when offering advice. \\
\midrule

Hobby \& Interest & HobbyDetails & Recall hobbies (e.g., photography) for related suggestions. \\
\addlinespace
Hobby \& Interest & OngoingProject & Refer to personal projects (e.g., building a model) when offering ideas. \\
\addlinespace
Hobby \& Interest & SeasonalActivityPreference & Recollect seasonal preferences when the season approaches. \\
\midrule

Shopping \& Purchase & PreferredBrands & Remember brand preferences for future suggestions. \\
\addlinespace
Shopping \& Purchase & PreviousPurchaseFeedback & Recall feedback on past purchases for similar recommendations. \\
\addlinespace
Shopping \& Purchase & PriceSensitivity & Keep budget/price preferences in mind for future recommendations. \\
\midrule

Emotional State & EmotionalState & Implicitly reference user's mood (e.g., stressed) in supportive suggestions. \\
\addlinespace
Emotional State & MentalHealthGoals & Recall goals (e.g., reducing anxiety) for wellness suggestions. \\
\addlinespace
Emotional State & RecentEmotionalExperience & Recollect recent experiences to offer empathy in future conversations. \\

\midrule
Behavioral Patterns & RecurringHesitation & Recall frequent hesitation points to offer reassurance. \\
\addlinespace
Behavioral Patterns & ProblemSolvingApproach & Remember preferred solution style when giving advice. \\
\addlinespace
Behavioral Patterns & ImplicitKnowledgeGap & Recall common knowledge gaps to clarify proactively. \\

% SELF COHERENCE
\midrule
\multicolumn{3}{c}{\textbf{\textsc{Self Coherence}}} \\
\midrule

Calculations & BudgetOrCostEstimates & Model must not provide conflicting financial figures later in the conversation. \\
\addlinespace
Calculations & TimeCalculations & Model must not give contradictory time estimates later. \\
\midrule

Factual Statements & HistoricalDates & Model must not contradict specific dates provided for historical events. \\
\addlinespace
Factual Statements & ScientificFacts & Model must ensure scientific facts are not contradicted in future statements. \\
\addlinespace
Factual Statements & GeographicalDetails & Model must not contradict geographical details (e.g., population) later. \\
\midrule

Recommendations & ProductRecommendations & Model must not contradict a product recommendation by suggesting something totally different later without cause. \\
\addlinespace
Recommendations & DietOrHealthAdvice & Model must avoid contradicting previously given dietary or health advice. \\
\addlinespace
Recommendations & TravelAdvice & Model must avoid contradictory travel suggestions. \\
\midrule

Contextualization & LanguageAndTone & Model must avoid contradictory shifts in tone (e.g., formal to friendly) unless prompted. \\
\midrule

Reasoning & CausalReasoning & Model must not contradict previously stated cause-and-effect explanations later. \\
\addlinespace
Reasoning & ArgumentativeCoherence & Model must not undermine earlier arguments with inconsistent logic later. \\
\addlinespace
Reasoning & ProConSynthesis & Model must not reverse previously stated trade-offs without justification. \\
\midrule

Plan & AdherenceToProposedPlan & Model must not forget details from a proposed plan. \\
\addlinespace
Plan Contradictions & GivenAdviceConsistency & Model must not give advice that conflicts with earlier guidance. \\
\midrule

Personas & RoleOrPersonaAdherence & Model must not break from an established role or persona mid-conversation. \\
\addlinespace
Personas & ScenarioConsistency & Model must not contradict established scenario assumptions later. \\
\addlinespace
Personas & AnalogyConsistency & Model must not change analogy mappings inconsistently once introduced. \\

\end{longtable}

\end{document}